\documentclass[8pt,a4paper]{jpconf}
\usepackage{mathrsfs}
\usepackage{amsmath,amssymb}
\usepackage[english, USenglish]{babel}

\newcommand{\beq}{\begin{equation}}
\newcommand{\eeq}{\end{equation}}
\newcommand{\beqa}{\begin{eqnarray}}
\newcommand{\eeqa}{\end{eqnarray}}
\newcommand{\nn}{\nonumber}

\newcommand{\mbb}{\mathbb}

\newcommand{\cZ}{\mathcal{Z}}
\newcommand{\cX}{\mathcal{X}}
\newcommand{\cM}{\mathcal{M}}
\newcommand{\cA}{\mathcal{A}}
\newcommand{\cB}{\mathcal{B}}
\newcommand{\cH}{\mathcal{H}}
\newcommand{\cN}{\mathcal{N}}
\newcommand{\cT}{\mathcal{T}}
\newcommand{\cL}{\mathcal{L}}
\newcommand{\cS}{\mathcal{S}}
\newcommand{\sF}{\mathscr{F}}

\newcommand{\sZ}{\mathscr{Z}}

\newcommand{\cC}{\mathcal{C}}
\newcommand{\sL}{\mathscr{L}}
\newcommand{\sH}{\mathscr{H}}
\newcommand{\cW}{\mathcal{W}}
\newcommand{\sE}{\mathscr{E}}
\newcommand{\cO}{\mathcal{O}}
\newcommand{\cR}{\mathcal{R}}
\newcommand{\cI}{\mathcal{I}}
\newcommand{\cE}{\mathcal{E}}
\newcommand{\nv}{n_{{}_{\text{V}}}}

\begin{document}
\title{Automorphic Instanton Partition Functions \\ on Calabi-Yau Threefolds}

\author{Daniel Persson\footnote{Also affiliated to {\it Fundamental Physics, Chalmers University of Technology, SE-412 96 Gothenburg, Sweden.}}}

\address{Instit\"ut f\"ur Theoretische Physik,  ETH Z\"urich, \\ CH-8093 Z\"urich, ÊSwitzerland}

\ead{daniel.persson@itp.phys.ethz.ch}

\vspace{.5cm}

\begin{abstract}
We survey recent results on quantum corrections to the hypermultiplet moduli space $\cM$ in type IIA/B string theory on a compact Calabi-Yau threefold $\cX$, or, equivalently, the vector multiplet moduli space in type IIB/A on $\cX\times S^{1}$. Our main focus lies on the problem of resumming the infinite series of D-brane and NS5-brane instantons, using the mathematical machinery of automorphic forms. We review the proposal that whenever the low-energy theory in $D=3$ exhibits an arithmetic ``U-duality'' symmetry $G(\mathbb{Z})$ the total instanton partition function arises from a certain unitary automorphic representation of $G$, whose Fourier coefficients reproduce the BPS-degeneracies.  For $D=4$, $\cN=2$ theories on $\mathbb{R}^3\times S^1$ we argue that the relevant automorphic representation falls in the quaternionic discrete series of $G$, and that the partition function can be realized as a holomorphic section on the twistor space $\cZ$ over $\cM$. We also offer some comments on the close relation with $\cN=2$ wall crossing formulae. \\

\noindent {\it Contribution to the proceedings of the workshop  on ``Algebra, Geometry and Mathematical Physics'' at the Tj\"arn\"o Marine Biological Laboratory, Sweden, October 2010. }

\end{abstract}

%\vspace{.1cm}
%\noindent {\footnotesize{\it }}

%\vspace{.5cm}

\section{Introduction}
U-duality symmetries \cite{Hull:1994ys} have provided remarkable insight into the non-perturbative nature of string theory and M-theory (see, e.g., \cite{Obers:1998fb}). The moduli space $\cM_4$ of string theory on $T^6$ is protected from quantum corrections and takes form of an arithmetic quotient $E_{7(7)}(\mathbb{Z})\backslash E_{7(7)} /(SU(8)/\mathbb{Z}_2)$ \cite{Hull:1994ys,Cremmer:1979up}. Moreover, higher derivative couplings in the resulting $\cN=8$ supergravity must be invariant under the U-duality group $E_{7(7)}(\mathbb{Z})$, and are thus naturally described within the mathematical theory of automorphic forms. This provides powerful methods of resumming infinite series of non-perturbative effects, such as D-brane and worldsheet instantons \cite{Green:1997tv,Obers:1999um,Lambert:2006ny,Bao:2007er,Bao:2007fx,Green:2010wi,Pioline:2010kb,Green:2010kv}. The Bekenstein-Hawking entropy of BPS black holes is furthermore invariant under U-duality \cite{Kallosh:1996uy}, and the associated microscopic degeneracies are given by Fourier coefficients of automorphic forms \cite{Maldacena:1997de,Dabholkar:2005dt,Pioline:2005vi}.  

The theory becomes even more constrained upon further compactification on $S^{1}$ to three dimensions. All dynamical degrees of freedom are then in the form of scalars, which parametrize an enlarged moduli space $E_{8(8)}(\mathbb{Z})\backslash E_{8(8)}/(Spin(16)/\mathbb{Z}_2)$, where the U-duality group is enhanced to $E_{8(8)}(\mathbb{Z})$. In fact, it is a general feature of four-dimensional supergravity theories with symmetry group $G_4$, that the enhanced group $G_3$ in three dimensions serves as a spectrum-generating symmetry of the original theory \cite{kinnersley,Breitenlohner:1987dg,Gunaydin:2000xr,Gunaydin:2001bt,Gunaydin:2005gd}. 

For $\cN=2$ supergravities the situation is much less under control. The moduli space is in general not a coset space, but is required by supersymmetry to factorize into a product $\cM_{{\text{V}}}\times \cM_{{\text{H}}}$, respectively corresponding to the vector-multiplet moduli space (special K\"ahler) and the hypermultiplet moduli space (quaternion-K\"ahler) \cite{Bagger:1983tt}. Moreover, the moduli space metric is no longer protected, and may receive both perturbative and non-perturbative quantum corrections. For theories arising from compactifications of string theory on a Calabi-Yau threefold $\cX$, there are perturbative corrections both in $\alpha'$ and in the string coupling $g_s=e^{\phi}$, together with non-perturbative effects arising from worldsheet instantons, Euclidean D-branes and NS5-branes wrapping (supersymmetric) cycles in $\cX$ \cite{Becker:1995kb} (for a review, see \cite{Aspinwall:2000fd,Pioline:2006ni}). 

One can also learn a lot about $\cN=2$ theories by compactification to $D=3$ \cite{Seiberg:1996nz}. In this case, the hypermultiplet moduli space goes along for the ride, while the vector multiplet moduli space is enhanced to another quaternion-K\"ahler manifold $\widehat{\cM_{\text{V}}}$; this procedure is known as the \emph{c-map} \cite{Cecotti:1988qn,Ferrara:1989ik}. The enhanced moduli space $\widehat{\cM_{\text{V}}}$ holds the key to understanding the wall-crossing behaviour of the BPS-index $\Omega(\gamma)$ (generalized Donaldson-Thomas invariants) \cite{ks,Joyce:2008pc,Gaiotto:2008cd,Alexandrov:2008gh}. In fact, using a combination of mirror symmetry and T-duality in three dimensions all the moduli spaces of type IIA  (resp. IIB) string theory on $\cX\times S^1$ (resp. $\hat{\cX}\times S^1$) become identified \cite{Becker:1995kb,Alexandrov:2008gh,Alexandrov:2009qq}. This unification of moduli has far-reaching consequences that goes beyond classical mirror symmetry \cite{Candelas:1990rm} as well as homological mirror symmetry \cite{MR1403918}. 

In this survey we shall discuss recent attempts at describing the exact moduli space $\widehat{\cM_{\text{V}}}$ in type IIA/B string theory on $\cX\times S^{1}$, or the hypermultiplet moduli space $\cM_{\text{H}}$ in type IIB/A on $\cX$. We argue that in certain classes of $\cN=2$ supergravities one can use automorphic techniques to constrain quantum corrections. In particular, having identified a ``U-duality'' group $G_3(\mathbb{Z})$, the BPS-degeneracies $\Omega(\gamma)$ are recovered from the Fourier coefficients of certain automorphic forms. Inspired by earlier related work \cite{Kazhdan:2001nx,Pioline:2005vi,Gunaydin:2005mx,Gunaydin:2006bz}, we revisit the recent results of \cite{Pioline:2009qt,Bao:2009fg,Bao:2010cc,Alexandrov:2010np,Alexandrov:2010ca} from a representation theoretic point of view, drawing insight from the ``toy models'' provided by so called \emph{magic supergravities} \cite{Gunaydin:1983rk,Gunaydin:1983bi}. We recall and elaborate upon conjectures stating that the instanton partition function in $\cN=2$ supergravity on $\mathbb{R}^{3}\times S^{1}$ should correspond to an automorphic representation in the quaternionic discrete series of  $G_3$. Along the way, we also make some comparative remarks on similarities with, and differences from, the better understood case of $\cN=2$ field theory  \cite{Gaiotto:2008cd}. 

%\vspace{.5cm}

%\noindent {\bf Notation.} We denote by $\cX$ a compact Calabi-Yau threefold, and by $\cX_{\text{r}}$ a CY threefold with $h_{2,1}(\cX_{\text{r}})=0$. The degree of supersymmetry $\cN$ is always measured in $D=4$. For $\cN\geq 4$ the 4d moduli space is denote by $\cM_4$ while $\cM_3$ is the enhanced moduli space in $D=3$. In  $\cN=2$ theories the complete moduli space is a product $\cM_{{}^{\text{V}}} \times \cM_{{}^{\text{H}}}$ of the vector multiplet and hypermultiplet moduli spaces, respectively. For type II on $\cX$, we sometimes refer to $\cM_{{}^{\text{V}}}$ as $\cM_{K}$ or $\cM_{C}$, respectively referring to the (complexified) K\"ahler or complex structure moduli of $\cX$. The enhanced vector multiplet moduli space in $D=3$ is denoted by $\widehat{\cM_{{}^{\text{V}}}}$. We denote by $t$ the collective coordinates on $\cM$, while $\{t, \zeta^{\Lambda}, \tilde\zeta_\Lambda, \sigma, \phi\}$ are coordinates on $\widehat{\cM_{{}^{\text{V}}}}$. $G_4$ and $G_3$ are real Lie groups, representing the continuous symmetries in $D=4$ and $D=3$, respectively, while $G_4(\mathbb{Z})$ and $G_3(\mathbb{Z})$ are discrete (U-duality) subgroups. For $\cN=8$, $G_4$ and $G_3$ are split real forms, while for $\cN=2$ they are in the ``quaternionic real form''. For CY-compactifications, $G_4(\mathbb{Z})$ is the monodromy group $\text{M}(\cX)$. 

%\newpage 
\section{Black Hole Partition Functions and Instanton Corrections}
\label{BHpartition}
In this section we introduce the physical arena in which our analysis takes place. We are mainly focusing on type IIA and IIB string compactifications on a compact Calabi-Yau threefold $\cX$. However, occasionally we will also be discussing other classes of $\cN=2$ supergravity theories, which do not necessarily have a known embedding in string theory. We will also be drawing some lessons from supergravities with a higher degree of supersymmetry. 

\subsection{BPS-States and Partition Functions in $\cN=2$ Theories}
\label{BPSstates}
Let us for definiteness begin with type IIB on $\cX$. The field content of the resulting four-dimensional $\cN=2$ theory splits between vectormultiplets and hypermultiplets. For discussing black holes, it suffices to restrict attention to the vectormultiplet sector, whose bosonic field content consists of the gravity multiplet, containing the 4d spacetime metric $g_{\mu\nu}$ and the graviphoton $A_\mu$, together with the vector multiplets, containing $n_{{}_{\text{V}}}=h_{2,1}(\cX)$ abelian vector fields $A_{\mu}^{I}$, and $h_{2,1}+1$ complex scalars $X^{\Lambda}$. The hypermultiplet sector will enter the story in sections \ref{Dbrane} and \ref{NS5analysis}.

We choose a Lagrangian splitting of $H_3(\cX, \mathbb{Z})=\Gamma_{\text{e}}\oplus \Gamma_{\text{m}}\equiv \Gamma$, with an associated symplectic basis $(\mathcal{A}^{\Lambda}, \cB_\Lambda),\, \Lambda=(0,I)$ of A- and B-cycles. The scalars $X^{\Lambda}$ then arise from the periods of the holomorphic 3-form $\Omega_{3,0}\in H^{3,0}(\mathcal{X}, \mathbb{C})$ along $\cA^{\Lambda}$, and thus parametrize the (special K\"ahler) complex structure moduli space $\cM_C(\cX)$. It is convenient to choose projective coordinates $t^{I}=X^{I}/X^{0}\in \cM_C(\cX)$, away from the locus $X^{0}=0$. 

The BPS-states arise from D3-branes wrapping special Lagrangian submanifolds (sLags) in the homology class $q_\Lambda \cA^{\Lambda}+p^{\Lambda}\cB_\Lambda \in H_3(\mathcal{X}, \mathbb{Z})$. We write the electric-magnetic charge vector as $\gamma=(q_\Lambda, p^{\Lambda})\in  \Gamma$. The mass $M$ of these states saturate the BPS-bound $M\geq |Z(\gamma; t)|$, where $Z(\gamma;t)\equiv \oint_\gamma \Omega_{3,0}$ is the central charge of the $\cN=2$ supersymmetry algebra. Mathematically, the central charge is a homomorphism $Z\, :\, \Gamma \rightarrow \mathbb{C}$, referred to as the ``stability data''  \cite{DouglasTopMirrorMap,MR2373143}. 

The black hole partition function $\sZ$ is a generating function of BPS-states. It is commonly attached to a ``mixed'' ensemble \cite{Ooguri:2004zv}, where the magnetic charges $p^{\Lambda}$ are held fixed, while summing over the electric charges $q_\Lambda$ weighted by conjugate chemical potentials $\zeta^{\Lambda}$. In general the partition function also varies over $\cM_C$, hence $\sZ=\sZ(p;t, \zeta)$. One then arrives at the following expression for the partition function in the mixed ensemble
\beq
\sZ_{\text{mix}}(p;t,\zeta)=\sum_{q_\Lambda\in  \Gamma_{\text{e}}} \Omega(q_\Lambda, p^{\Lambda};t) e^{2\pi i q_\Lambda \zeta^\Lambda},
\eeq
where $\Omega : \Gamma\, \rightarrow \, \mathbb{Z}$ is the BPS-index which counts the weighted number of BPS-states of charge vector $\gamma$. Physically, the index corresponds to the ``second helicity supertrace'' which is defined as a trace over the BPS Hilbert space $\cH_{\gamma,t}^{{}_{\text{BPS}}}$ \cite{Kiritsis:1997gu,Dabholkar:2005dt}: 
\beq
\Omega(\gamma;t)=\frac{1}{2}\text{Tr}_{\cH_{\gamma, t}^{{}_{\text{BPS}}}} (2J_3)^{2}(-1)^{2J_3},
\label{secondhelicity}
\eeq
where $J_3$ is a generator of the little group $SO(3)$ in four dimensions. The insertion of $(2J_3)^2$ is required to obtain a non-vanishing result in the presence of $\cN=2$ supersymmetry. The BPS-index conjecturally counts objects in the derived Fukaya category of $\cX$, and by mirror symmetry should be equivalent to the generalized Donaldson-Thomas invariants \cite{MR1403918} (see \cite{Aspinwall:2004jr} for a review). The index $\Omega(\gamma;t)$ is furthermore a locally constant function of $t$, but may jump for special values corresponding to real codimension one submanifolds of $\cM_C$, known as walls of marginal stability \cite{Denef:2007vg,Gaiotto:2008cd,Joyce:2008pc,ks,Alexandrov:2008gh,Alexandrov:2009zh}. 

For our present purposes it is natural to consider a slightly different generating function, namely one which is covariant with respect to symplectic transformations on the electric-magnetic charges. This corresponds to a partition function attached to a ``canonical ensemble'' \cite{LopesCardoso:2006bg}, where one introduces chemical potentials $\tilde\zeta_\Lambda$ also for the magnetic charges $p^{\Lambda}$ and the sum is taken over the full charge lattice. We define such a canonical partition function as
\beq 
\sZ_{\text{can}}(t,\zeta, \tilde\zeta)=\sum_{\gamma\in \Gamma }\lambda(\gamma)\overline{\Omega}(\gamma;t)e^{2\pi i (q_\Lambda\zeta^{\Lambda}-p^{\Lambda}\tilde\zeta_\Lambda)},
\label{canonicalpartitionfunction}
\eeq
where we have replaced the integral degeneracies $\Omega(\gamma)\in \mathbb{Z}$ with their rational counterparts $\overline{\Omega}(\gamma)\in \mathbb{Q}$, which are related as follows:
\beq
\overline{\Omega}(\gamma)=\sum_{d|\gamma}\frac{\Omega(\gamma/d)}{d^2}, \qquad \qquad \Omega(\gamma)=\sum_{d|\gamma}\frac{\mu(d)\overline{\Omega}(\gamma/d)}{d^2},
\eeq
with $\mu(d)$ being the M\"obius function. The rational degeneracies $\overline{\Omega}(\gamma)$ coincide with the Donaldson-Thomas  invariants introduced by Joyce and Song \cite{Joyce:2008pc}, and are more suited for our purposes. In particular, they are better behaved under wall crossing and S-duality \cite{Manschot:2010xp,Alexandrov:2010ca,Manschot:2010qz}. In (\ref{canonicalpartitionfunction}) we have also inserted a $U(1)$-valued function $\lambda(\gamma)$ which corresponds to a quadratic refinement of the symplectic pairing on $\Gamma$. It can be viewed as a generalization of the sign $(-1)^{p^{\Lambda}q_\Lambda}$ familiar from D0-D2-D4 partition functions \cite{Gaiotto:2006wm,Manschot:2009ia,Manschot:2010xp}, and is ultimately required for describing the wall crossing behaviour of $\overline{\Omega}(\gamma)$ \cite{Gaiotto:2008cd}. We will discuss properties of $\lambda(\gamma)$ in more detail in section \ref{NS5analysis}.

\subsection{Instanton Partition Functions in Three Dimensions}
\label{3dinstanton}

We now wish to consider the theory on $\mathbb{R}^3\times S^{1}$ and analyze the low-energy physics in $D=3$. Part of our motivation comes from string compactifications preserving $\cN\geq 4$ supersymmetries, for which it is well-known that the 4d effective action is invariant under a discrete U-duality group $G_4(\mathbb{Z})$. In these cases the moduli space $\cM_4$ of vacua Êis protected from quantum corrections, and takes the form of an arithmetic coset space $\cM_4=G_4(\mbb{Z})\backslash G_4/K_4$, where $G_4$ is the continuous symmetry of the classical theory, and $K_4$ is the maximal compact subgroup of $G_4$, playing the role of a local ``R-symmetry''. For torus compactifications the symmetry group is $E_{7(7)}(\mbb{Z})$ \cite{Hull:1994ys}, while for $K3\times T^2$ it is $SL(2,\mathbb{Z})\times SO(6, 22;\mathbb{Z})$. By further compactification on a circle of radius $R=e^{-\phi}\ell_{\text{P}}$ ($\ell_{\text{P}}$ being the 4d Planck length), the U-duality group gets enhanced to a larger group $G_3(\mbb{Z})\supset G_4(\mbb{Z})$. Moreover, in three dimensions all $p$-form degrees of freedom can be dualized to (pseudo-)scalars which live in the enlarged moduli space\footnote{If the circle $S^1$ is timelike one obtains instead a pseudo-Riemannian coset space $\cM_3^{*}=G_3/K_3^{*}$, where $K_3^{*}$ Êis non-compact, whose non-positive definite metric can be obtained by analytic continuation from $\cM_3$ \cite{Breitenlohner:1987dg}. We will not distinguish between $\cM_3$ and $\cM_3^{*}$ in the following. } 
\beq 
\cM_3=G_3(\mbb{Z})\backslash G_3/K_3.
\eeq
Invariance under the spectrum generating symmetry $G_3(\mathbb{Z})$ provides powerful constraints on quantum corrections in the effective action; in particular, BPS-saturated higher derivative couplings are $G_3(\mathbb{Z})$-invariant Eisenstein series (see, e.g., \cite{Obers:1999um}).

Let us now perform a similar analysis in the $\cN=2$ setting. For definiteness, we still restrict to the type IIB picture. The role of the symmetry group $G_4(\mathbb{Z})$ is played by the monodromy group $\text{M}(\cX)\subset Sp(2h_{2,1}+2;\mathbb{Z})$ of the Calabi-Yau threefold $\cX$. After compactification to $D=3$, one finds two new moduli from the gravity multiplet: the radius $e^{-\phi}$ of the circle, together with the ``NUT potential'' $\sigma$, corresponding to the dual of the Kaluza-Klein vector (the off-diagonal component of the 4d metric).   In addition, there are $2(\nv+1)$ scalars arising from the electric and magnetic Wilson lines of the $\nv+1=h_{2,1}+1$ abelian vector fields $A_\mu^{\Lambda}$ (including the graviphoton) along the circle:
\beq
\zeta^{\Lambda} = \oint_{S^1} A_R^{\Lambda} dR,\qquad \qquad \tilde\zeta_\Lambda = \oint_{S^1} (A_R^{\star})_\Lambda dR,
\label{Wilson}
\eeq
where $A_R^{\Lambda}$ is the component of $A_\mu^{\Lambda}$ along the circle, while $(A^{\star}_R)_\Lambda$ is the same component of the dual vector field in $D=4$. Since Wilson lines couple to electric (or magnetic) charges $Q$ through terms of the form $e^{2\pi iQ\oint_R A dR}$, we may, after comparison with (\ref{canonicalpartitionfunction}), identify the Wilson lines with the chemical potentials, thereby explaining our choice of notation in (\ref{Wilson}). Thus, one of the virtues of compactification to three dimensions is that the auxiliary chemical potentials $(\zeta^{\Lambda}, \tilde\zeta_\Lambda)$ become promoted to actual physical degrees of freedom. Including also the complex structure moduli $t\in \cM_C(\cX)$, the scalars $\{t, \phi, \zeta^{\Lambda}, \tilde\zeta_\Lambda, \sigma\}$ parametrize the enhanced vector multiplet moduli space in $D=3$, which we shall denote by $\widehat{\cM_C}(\cX)$. The moduli space $\widehat{\cM_C}(\cX)$ is real $4(h_{2,1}+1)$-dimensional and is equipped with a \emph{quaternion-K\"ahler} metric \cite{Bagger:1983tt,Cecotti:1988qn,Ferrara:1989ik}.\footnote{Upon reduction on a timelike circle $S^{1}$ the resulting moduli space $\widehat{\cM^{*}_C}(\cX)$ is ``para-quaternion-K\"ahler'' \cite{Cortes:2005uq}, which is the quaternionic analogue of the pseudo-Riemannian coset space $\cM_3^{*}$.}   

In contrast to $\cM_3$, $\widehat{\cM_C}(\cX)$ is in general not a coset space, and is also not protected from quantum corrections. The classical, or ``semiflat'' \cite{Gaiotto:2008cd}, metric $\widehat{g}^{\, \text{sf}}$ on $\widehat{\cM_C}(\cX)$ is obtained by naive dimensional reduction on $S^1$. This metric is singular but it is expected that quantum corrections ensure that the exact moduli space metric $\widehat{g}$ is smooth. These corrections are due to instanton effects which arise from $D=4$ Euclidean BPS-states wrapping the circle \cite{Seiberg:1996nz}. Microscopically, they correspond to  Euclidean D3-branes wrapping $\cL\times S^{1}$, where $\cL\subset \cX$ is a sLag. The quantum deformations of the semiflat metric are weighted by the exponential of minus the radius of the circle times the mass of the BPS-state (or black hole), schematically
\beq
\widehat{g}\, \sim\,  \widehat{g}^{\, \text{sf}} +e^{-2\pi e^{-\phi}|Z(\gamma, t)|}  .
\label{semiflat}
\eeq
The explicit form of $\widehat{g}^{\, \text{sf}}$ Êis given in eq. (\ref{semiflat}) below. The instanton corrections are exponentially suppressed in the large radius limit $e^{-\phi}\rightarrow \infty$, but contributes with subleading effects for finite values of $e^{-\phi}$. From (\ref{Wilson}) we learn that the Wilson lines couple to the charges $\gamma=(q_\Lambda, p^{\Lambda})$ via terms of the form $e^{-2\pi i (q_\Lambda \zeta^{\Lambda}-p^{\Lambda}\tilde\zeta_\Lambda)}$, and hence multi-instanton effects are captured by the partition function 
\beq
\sZ_{{}^{\text{BPS}}}(t, \phi, \zeta, \tilde\zeta)\sim \sum_{\gamma\in \Gamma}\lambda(\gamma) \overline{\Omega}(\gamma; t) e^{-2\pi \cS^{{}_\text{BPS}}_\gamma( t,R, \zeta, \tilde\zeta)},
\label{BPSinst}
\eeq
where $\cS^{{}_{\text{BPS}}}_\gamma$ is the Euclidean action for a BPS-instanton of charge $\gamma$:
\beq
\cS^{{}_\text{BPS}}_\gamma(t,R, \zeta, \tilde\zeta)=e^{-\phi}|Z(\gamma;t)| + i(q_\Lambda\zeta^{\Lambda}-p^{\Lambda}\tilde\zeta_\Lambda).
\eeq
$\sZ_{{}^{\text{BPS}}}$ should be viewed as the instanton analogue of the black hole partition function (\ref{canonicalpartitionfunction}). Charge quantization (``large gauge transformations'') requires that $(\zeta^{\Lambda}, \tilde\zeta_\Lambda)$ are angle-valued:
\beq
\zeta^{\Lambda} \longmapsto \zeta^{\Lambda} + m^{\Lambda}, \qquad \tilde\zeta_\Lambda \longmapsto \tilde\zeta_\Lambda +n_\Lambda, \label{torusshift}
\eeq
where $(m^{\Lambda}, n_\Lambda)\in  \mathbb{Z}^{2(\nv+1)}$. Hence, $(\zeta^{\Lambda}, \tilde\zeta_\Lambda)$ parametrize a symplectic torus $\cT(\cX)=T^{2(\nv+1)}$.

Note that the BPS-index $\overline{\Omega}(\gamma;t)$ reappeared in (\ref{BPSinst}) as the ``instanton measure'' which encodes multiplicities of instanton bound states in three dimensions. Physically, it corresponds to the path integral measure on the instanton moduli space \cite{Green:1997tn}. The assumption that  these two numerical quantities are equal is strongly supported by the behaviour of $\overline{\Omega}(\gamma;t)$ under wall crossing \cite{Gaiotto:2008cd} (see \cite{Yi:1997eg,Sethi:1997pa,Green:1997tn,Kostov:1998pg,Moore:1998et,Alexandrov:2008gh,Pioline:2009ia} for related discussions).
 
A crucial difference between $\cN=2$ field theory and $\cN=2$ supergravity is that for large charges the indexed degeneracies of BPS states grow exponentially
\beq
\overline{\Omega}(\gamma)\sim \exp\big[ S_{{}^\text{BH}}(\gamma)\big], 
\label{Fouriergrowth}
\eeq
where $S_{{}^\text{BH}}(\gamma)$ is the Bekenstein-Hawking entropy of a black hole with charge $\gamma$. This implies that the instanton sum in (\ref{BPSinst}) diverges, in contrast to the analogous sum in field theory \cite{Gaiotto:2008cd}. A possible resolution to this problem was suggested in \cite{Pioline:2009ia}, by interpreting (\ref{BPSinst}) as an asymptotic series, the ambiguity of which is of the order $\exp[-e^{-2\phi}]$. Such effects indeed exist in $D=3$, and are precisely due to the presence of gravity, as we will now discuss. 

In addition to the BPS-instantons, the semiflat metric (\ref{semiflat}) also receives corrections from gravitational instantons which have no physical analogue in $D=4$ \cite{Behrndt:1997ch,Pioline:2005vi,Pioline:2006ni,Gunaydin:2007bg,Alexandrov:2008gh}. More precisely, they correspond to four-dimensional Taub-NUT solutions with closed timelike curves. However, in $D=3$ these give rise to physical non-perturbative effects, whose contributions to the metric on $\widehat{\cM_C}(\cX)$ are of the form $e^{-2\pi |k| e^{-2\phi}}$ and are therefore subleading in the large radius limit. The associated gravitational scalar $\sigma$ couples to the NUT charge $k\in \mathbb{Z}$ through terms of the form $e^{-i\pi k \sigma}$. This requires that $\sigma$ is periodic modulo 2 (in our conventions), and furthermore that it picks up non-trivial shifts under translations along the torus $\cT(\cX)$ \cite{Alexandrov:2010np,Alexandrov:2010ca}:
\beq
\sigma \longmapsto \sigma + 2r -n_\Lambda \zeta^{\Lambda} + m^{\Lambda}\tilde\zeta_\Lambda+2\upsilon(m,  n),
\label{Heis}
\eeq
where $r\in \mathbb{Z}$. Here we have allowed for the presence of a cocycle shift $\upsilon(m,  n)$ whose appearance is related to the quadratic refinement $\lambda(\gamma)$ as will be explained in section \ref{NS5analysis}. 

We conclude that the semiflat metric on $\widehat{\cM_C}$ is also deformed by gravitational  instantons captured by a partition function of the form
\beq
\sZ_{{}^{\text{TN}}}(t, \phi, \zeta, \tilde\zeta, \sigma) \sim \sum_{k\neq 0} \sZ^{(k)}(t, \zeta, \tilde\zeta) e^{-2\pi \cS_k^{{}_\text{TN}}(\phi, \sigma)},
\label{TaubNutInst}
\eeq
where the Euclidean action of a charge $k$ Taub-NUT instanton is
\beq
\cS_k^{{}_\text{TN}}(\phi, \sigma)= |k| e^{-2\phi} + \tfrac{1}{2} ik \sigma.
\eeq
We defer a discussion of the coefficients $\sZ^{(k)}(t, \zeta, \tilde\zeta) $ to section \ref{NS5analysis}. 

Combining the contributions in (\ref{BPSinst}) and (\ref{TaubNutInst}) we find that the exact metric on the moduli space $\widehat{\cM_C}$ 
should take the schematic form
\beq
\widehat{g}\, = \,  \widehat{g}^{\, \text{sf}} +\sZ_{{}^{\text{tot}}}(t, \phi, \zeta, \tilde\zeta, \sigma)  ,
\label{semiflat}
\eeq
where $\sZ_{{}^{\text{tot}}}$ is the total instanton partition function whose large-radius expansion is given by 
\beq
\sZ_{{}^{\text{tot}}}(t, \phi, \zeta, \tilde\zeta, \sigma) \sim  \sum_{\gamma\in \Gamma} \lambda(\gamma) \overline{\Omega}(\gamma; t) e^{-2\pi \cS^{{}_\text{BPS}}_\gamma(t, \phi, \zeta, \tilde\zeta)}+\sum_{k\neq 0} \sZ^{(k)}(t, \zeta, \tilde\zeta)e^{-2\pi \cS_k^{{}_\text{TN}}(\phi, \sigma)} .
\label{totalinstantoncontribution}
\eeq

\subsection{Topology of the Moduli Space and the Semiflat Metric}

Here we offer some useful remarks on the topology of the moduli space $\widehat{\cM_C}(\cX)$, which will in particular allow for a nice explicit  presentation of the semiflat metric $\widehat{g}^{\, \text{sf}}$. By virtue of eq. (\ref{Heis}), the shifts of $(\zeta^{\Lambda}, \tilde\zeta_\Lambda)$ do not commute, and the total translation group acting on $(\zeta^{\Lambda}, \tilde\zeta_\Lambda, \sigma)$ is identified with a $(2\nv+3)$-dimensional arithmetic Heisenberg group $N(\mbb{Z})$. This implies that the coordinates $(\zeta^{\Lambda}, \tilde\zeta_\Lambda, \sigma)$ parametrize the total space of a circle bundle $\cC_\sigma \rightarrow \cT(\cX)$, with fiber given by the circle $S^{1}_\sigma$ parametrized by $\sigma$. Moreover, the base $\cT(\cX)$ is itself non-trivially fibered over the complex structure moduli space $\cM_C(\cX)$, and we denote the total space of this torus fibration by $\mathscr{J}_{C}(\cX)$:
\beq
\cT(\cX)\quad \longrightarrow \quad \mathscr{J}_{C}(\cX) \quad \longrightarrow \quad \cM_C(\cX) \, .
\label{torusfibration}
\eeq
At fixed (large) value of the radius $R>> \ell_{\text{P}}$ $(e^{-\phi}\rightarrow \infty)$, the three-dimensional moduli space then exhibits a foliation \cite{Alexandrov:2010np,Alexandrov:2010ca}
\beq
\widehat{\cM_C}(\cX)\sim \mathbb{R}_+\times \cC_\sigma(\phi)
\label{foliation}
\eeq
where the leaves are circle bundles $\cC_\sigma(\phi)$ over $\mathscr{J}_{C}(\cX)$: 
\beq
S^{1}_{\sigma}\quad   \longrightarrow\quad   \cC_\sigma(\phi)\quad  \longrightarrow\quad  \mathscr{J}_{\cM_4} \, .
\label{3Dfibration}
\eeq 
This represents a supergravity generalization of the Seiberg-Witten  (hyperk\"ahler) torus fibration over the Coulomb branch $\cB$ of $\cN=2$ field theories on $\mathbb{R}^3\times S^1$ \cite{Seiberg:1996nz,Gaiotto:2008cd}. Having identified the topology of $\widehat{\cM_C}(\cX)$ through the foliation (\ref{foliation}), we can now write the semiflat metric $\widehat{g}^{\, \text{sf}}$ on $\widehat{\cM_C}(\cX)$ in the following explicit form\footnote{See \cite{Cecotti:1988qn,Ferrara:1989ik} for the original form of this metric, where it was referred to as the \emph{c-map metric}.}
\beq
\widehat{g}^{\, \text{sf}}= 4\text{d}\phi^2+4 g_{\cM_C}+e^{2\phi} g_{\cT}+\tfrac{1}{16} e^{4\phi} \cA_{\cC_\sigma}^2,
\label{semiflat}
\eeq
where $\cA_{\cC_\sigma}=\text{d}\sigma+\tilde\zeta_\Lambda\text{d}\zeta^{\Lambda}-\zeta^{\Lambda}\text{d}\tilde\zeta_\Lambda$ is the connection on the circle bundle $\cC_\sigma(\phi)$, with first Chern class $c_1(\cC_\sigma)=\text{d}(\cA_{\cC_\sigma}/2)=\text{d}\tilde{\zeta}_\Lambda\wedge \text{d}\zeta^{\Lambda}$ equal to the K\"ahler form $\omega_\cT$ on the torus $\cT(\cX)$ \cite{Alexandrov:2010ca}.\footnote{In string theory there is also a one-loop correction to the semiflat metric \cite{Antoniadis:1997eg,Gunther:1998sc,Antoniadis:2003sw}, which modifies the topology of $\cC_\sigma$ \cite{Alexandrov:2010ca}, although this will not concern us in the present work.}

%The ultimate goal of our analysis is to resum the series in (\ref{totalinstantoncontribution}) to find an expression for exact instanton partition function $\sZ_{{}^{\text{tot}}} $. As guidance, we first recall that there is a $(2\nv+3)$-dimensional discrete Heisenberg group $N(\mathbb{Z})$ acting on $(\zeta^{\Lambda}, \tilde\zeta_\Lambda, \sigma)$, under which the partition function (\ref{totalinstantoncontribution}) must remain invariant. In addition, the metric on $\cM_4$ carries an isometric action of the 4d duality group $G_4(\mathbb{Z})$, which for Calabi-Yau compactifications is the monodromy group $\text{M}(\cX)$. We now want to entertain the possibility that that the semi-direct product $G_4(\mathbb{Z})\ltimes N(\mathbb{Z})$ fits into a larger U-duality symmetry $G_3(\mathbb{Z})$ which acts isometrically on $\widehat{\cM_C}(\cX)$ and leaves $\sZ_{{}^{\text{tot}}} $ invariant. The consequences of this will be elaborated upon in section \ref{instpart}, after we have gained more insight into (\ref{totalinstantoncontribution}) using a combination of T-duality and mirror symmetry. 

\subsection{T-Duality and D-Brane Instantons}
\label{Dbrane}
We now perform a T-duality transformation along the circle $S_\phi^{1}$, which maps the moduli space $\widehat{\cM_C}(\cX)$ into the hypermultiplet moduli space $\cM_{{}^\text{H}}^{{}_{\text{IIA}}}(\cX)$ of type IIA string theory  on $\cX $. The radius $e^{-\phi}$ of the circle is then mapped to the four-dimensional dilaton, $g_s\equiv e^{\phi}$, in such a way that the large radius limit in the type IIB picture is equivalent to the weak-coupling limit $g_s \rightarrow 0$ on the type IIA side. After T-duality the Wilson lines $(\zeta^{\Lambda}, \tilde\zeta_\Lambda)$ correspond to the periods of the type IIA Ramond-Ramond 3-form $C_{(3)}$ over a symplectic basis of $H_3(\cX, \mathbb{Z})$:
\beq
\zeta^{\Lambda}=\int_{\cA^{\Lambda}}C_{(3)}, \qquad \qquad \tilde\zeta_\Lambda=\int_{\cB_\Lambda} C_{(3)}, 
\eeq
and the torus $\cT(\cX)$ is thereby identified with the intermediate Jacobian, 
\beq 
\cT(\cX)=H^{3}(\cX, \mbb{R})/H^{3}(\cX, \mathbb{Z}). 
\eeq
The torus fibration (\ref{torusfibration}) thus encodes the fact that the complex structure of $\cT(\cX)$ varies over $\cM_C(\cX)$ and the total space $\mathscr{J}_C(\cX)\rightarrow \cM_C(\cX)$ represents the entire family of intermediate Jacobians of $\cX$. Since $C_{(3)}$ couples to the D2-brane, we conclude that the instanton effects in (\ref{BPSinst}) are T-dual to Euclidean D2-branes wrapping sLags $\cL\subset \cX$. Indeed, this is consistent with the fact that in the weak-coupling limit, D-instantons are exponentially suppressed by $e^{-1/g_s}$.

\subsection{T-Duality, NS5-Branes and Generalized Theta Functions}
\label{NS5analysis}

On the other hand, after T-duality the Taub-NUT instantons are mapped into non-perturbative effects arising from Euclidean NS5-branes wrapping the Calabi-Yau manifold $\cX$ \cite{Ooguri:1996wj,Kapustin:2004jm,Dijkgraaf:2002ac}. The form of the coupling (\ref{TaubNutInst}) is consistent with the fact that NS5-brane effects are weighted by $e^{-1/g_s^{2}}$ \cite{Becker:1995kb}. The scalar $\sigma$ corresponds to the NS-axion, arising from the dualization of the Kalb-Ramond 2-form $B_{(2)}$ in $D=4$. 

These observations provide information about the structure of the gravitational partition function (\ref{TaubNutInst}). Invoking the results of \cite{Alexandrov:2010np,Alexandrov:2010ca} we deduce that the coefficient $\mathscr{Z}^{(k)}$ corresponds to the partition function of a charge $k$Ê Euclidean NS5-brane wrapped on the Calabi-Yau manifold $\cX$. Since the worldvolume theory of the type IIA NS5 brane is chiral \cite{Callan:1991ky}, $\mathscr{Z}^{(k)}$ is not a proper function, but rather a holomorphic section of a line bundle $\mathscr{L}^{k}$ over $\mathscr{J}_C(\cX)$ \cite{Witten:1996hc,Henningson:1999dm,Belov:2006jd}. 

These results imply that $\sZ^{(k)}$ satisfies twisted periodicity relations under translations by a vector $(m^{\Lambda},  n_\Lambda)\in H^{3}(\cX, \mathbb{Z})$ along the torus $\cT$:
\beq
\sZ^{(k)}(t, \zeta+m, \tilde\zeta+ n)=[\lambda(m,  n)]^{k} e^{\pi ik(m^{\Lambda}\tilde\zeta_\Lambda- n_\Lambda\zeta^{\Lambda})} \sZ^{(k)}(t, \zeta, \tilde\zeta),
\label{twistedperiodicity}
\eeq
where $\lambda(m,  n)$ is a homomorphism $\lambda : H^{3}(\cX, \mathbb{Z}) \rightarrow U(1)$, corresponding to a quadratic refinement of the intersection form on $H_3(\cX, \mathbb{Z})$ \cite{Witten:1996hc,Belov:2006jd}. It is defined by the cocycle relation:
\beq
\lambda(H+H')=(-1)^{\left<H,H'\right>}\lambda(H)\lambda(H'),
\label{cocycle}
\eeq
where $H\in H^{3}(\cX, \mathbb{Z})$. Equivalently, $\lambda$ describes the holonomies around one-cycles in $\cT(\cX)$. For the choice of Lagrangian decomposition of $H^{3}(\cX, \mathbb{Z})$ corresponding to $H=(m^{\Lambda},  n_\Lambda)$, one can solve the condition (\ref{cocycle}) generally in terms of characteristics $\Theta=(\theta^{\Lambda}, \phi_\Lambda)$ such that \cite{Belov:2006jd}
\beq
\lambda(m,  n)=e^{-i\pi  n_\Lambda m^{\Lambda}+2\pi i ( n_\Lambda\theta^{\Lambda}-m^{\Lambda}\phi_\Lambda)}\equiv (-1)^{2\upsilon(m,  n)},
\eeq
where $\upsilon(m,  n)$ is defined modulo one. It was conjectured in \cite{Alexandrov:2010np,Alexandrov:2010ca}---based on S-duality and other considerations---that the quadratic refinement $\lambda(m, n)$ governing the NS5-partition function is equivalent to the $\lambda(\gamma)$ appearing in (\ref{canonicalpartitionfunction}), after replacing the ``flux'' $H=(m^{\Lambda}, n_\Lambda)$ in its argument by the electric-magnetic charge vector $\gamma=(p^{\Lambda}, q_\Lambda)$. 

Sections of $\sL$ can be represented by generalized theta series of the form \cite{Alexandrov:2010np}
\beq
\sZ^{(1)}_{ \Theta}(t, \zeta, \tilde\zeta)=\sum_{p^{\Lambda}\in \Gamma_{\text{m}}+\theta^{\Lambda}} \Psi(t, \zeta^{\Lambda}-p^{\Lambda}) e^{2\pi i (\tilde\zeta_\Lambda-\phi_\Lambda)p^{\Lambda}+i\pi (\theta^{\Lambda}\phi_\Lambda-\zeta^{\Lambda}\tilde\zeta_\Lambda)},
\eeq
where $\Psi(\zeta^{\Lambda})$ is a ``wave function'' on $\Gamma_{\text{m}}\otimes \mathbb{R}$. When $\Psi(\zeta^{\Lambda})$ is a simple Gaussian then $\sZ^{(1)}_{\Theta}$ is a holomorphic section of $\sL$ and coincides with the partition function of the self-dual field, as analyzed in \cite{Witten:1996hc,Belov:2006jd}. For $k>1$, the bundle $\sL^{k}$ admits $|k|^{2h_{2,1}(\cX)}$ holomorphic sections corresponding to the elements in $\Gamma_{\text{m}}/(|k|\Gamma_{\text{m}})$, and the total partition function becomes a sum of vector-valued theta series \cite{Alexandrov:2010ca}:
\beqa
\sZ_\Theta^{(k)}(t, \zeta, \tilde\zeta)&=&\sum_{\ell^{\Lambda}\in \Gamma_{\text{m}}/(|k|\Gamma_{\text{m}})} \sZ^{k, \ell^{\Lambda}}_\Theta(t, \zeta, \tilde\zeta)
\nn \\
&=&\sum_{\ell^{\Lambda}\in \Gamma_{\text{m}}/(|k|\Gamma_{\text{m}})}\sum_{p^{\Lambda}\in \Gamma_{\text{m}}+\theta^{\Lambda}+\ell^{\Lambda}} \Psi_{k, \ell^{\Lambda}}(t, \zeta^{\Lambda}-p^{\Lambda}) e^{2\pi i k (\tilde\zeta_\Lambda-\phi_\Lambda)p^{\Lambda}+i\pi k (\theta^{\Lambda}\phi_\Lambda-\zeta^{\Lambda}\tilde\zeta_\Lambda)}.
\nn \\
\label{vectorvaluedtheta}
\eeqa

By examining the coupling (\ref{TaubNutInst}) we deduce that the Heisenberg transformation of $\sigma$ induced by translations of $(\zeta^{\Lambda}, \tilde\zeta_\Lambda)$ is compensated by the middle term in the transformation  (\ref{twistedperiodicity}) of the partition function $\sZ^{(k)}_{ \Theta}$. However, it follows from (\ref{twistedperiodicity}) that this compensating transformation also receives a contribution $[\lambda(m,  n)]^{k}$ from the quadratic refinement. This implies that for the sum (\ref{TaubNutInst}) to be well-defined from the point of view of the line bundle $\sL$, the Heisenberg transformation of the NS-axion $\sigma$ must   contain a term involving the cocycle $\upsilon(m,n)$ \cite{Alexandrov:2010np,Alexandrov:2010ca}, as was already anticipated in eq. (\ref{Heis}). In other words, $e^{i\pi \sigma}$ is valued in the same line bundle $\sL$ as the NS5 partition function $\sZ^{(k)}_{ \Theta}$.

%\subsection{Monodromies and the Metaplectic Representation}

In the general non-linear case when the wave function $\Psi_{k, \ell^{\Lambda}}(t, \zeta^{\Lambda}) $ is non-Gaussian, the partition function  $\sZ^{(k)}_{ \Theta}$ is no longer a holomorphic section of $\sL^k$. In fact, it was shown in \cite{Alexandrov:2010ca} that S-duality and mirror symmetry implies that for $k=1$ the wave function $ \Psi_{1,0}(\zeta^{\Lambda})$ should be identified with the real-polarized partition function $\Psi_{\text{top}}=\exp\big(\sum_g \lambda^{2g-2} F_g\big)$ of the topological string on $\cX$, consistently with earlier related results \cite{Kapustin:2004jm,Nekrasov:2004js,Dijkgraaf:2002ac,Bao:2006ef,Bengtsson:2008sp}. These observations have bearing on the transformation properties of the partition function $\sZ_\Theta^{(k)}$ under the monodromy group $\text{M}(\cX)$, or, more generally, under the 4d Êduality group $G_4(\mathbb{Z})$. In particular, the wave function $\Psi_{k, \ell^{\Lambda}}(\zeta^{\Lambda})$ transforms in the Schr\"odinger-Weil representation of $G_4(\mathbb{Z})\ltimes N(\mathbb{Z})$ \cite{Gunaydin:2006bz,Alexandrov:2010ca}.\footnote{The precise dependence of the NS5-partition function on the Calabi-Yau metric, i.e. on the moduli $t\in \cM_C$, is subtle \cite{Moore:2004jv,Belov:2006jd,Monnier:2010ww}, and is closely related to the one-loop correction to the semiflat metric $\widehat{g}^{\, \text{sf}}$ \cite{Alexandrov:2010np,Alexandrov:2010ca}.}

\subsection{Mirror Symmetry and Unification of Moduli in $D=3$}

Let us now also briefly discuss the mirror version of the above story. Mirror symmetry equates the complex structure moduli space $\cM_C(\cX)$ with the complexified K\"ahler moduli space $\cM_K(\hat{\cX})$ of the mirror Calabi-Yau threefold $\hat{\cX}$. Physically, $\cM_{K}(\hat{\cX})$ is the vector multiplet moduli space of type IIA string theory compactified on $\hat{\cX}$. Under mirror symmetry, the spectrum of BPS-states states arising from D3-branes wrapping special Lagrangian submanifolds $\cL\subset \cX$, is mapped to bound states of D0-D2-D4-D6 branes wrapping even cycles in the Calabi-Yau. Such a bound state is accurately described by a stable coherent sheaf $\cE$ on $\hat{\cX}$, with charges classified by the K-theory group $K(\hat{\cX})$ \cite{Minasian:1997mm}. Concretely, the charges are described through the Mukai vector $\gamma': K(\hat{\cX}) \rightarrow H_{\text{even}}(\hat{\cX}, \mathbb{Q})$, which takes the explicit form 
\beq
\gamma'=\text{ch}(\cE)\sqrt{\text{Td}(\hat{\cX})},
\eeq
where $\text{ch}(\cE)$ Êis the Chern character of the sheaf $\cE$ and $\text{Td}(\hat{\cX})$ is the Todd class of $T\hat{\cX}$. Since $\gamma'$ is valued in the rational cohomology group $H_{\text{even}}(\hat{\cX}, \mathbb{Q})$ one cannot immediately identify it with its mirror dual $\gamma\in H_3(\cX, \mathbb{Z})$. To properly identify the charges under mirror symmetry one must make a symplectic transformation on the electric charges
\beq
A\, :\,  \mathbb{Q}\, \ni \, q_\Lambda' \, \mapsto \, q_\Lambda\equiv q_\Lambda'+A_{\Lambda\Sigma}p^{\Sigma} \, \in \mathbb{Z}, 
\eeq
where $A_{\Lambda\Sigma}$ is a certain integer-valued symplectic matrix (see \cite{Alexandrov:2010ca} for more details). Thus, mirror symmetry relates $\gamma\in H_{3}(\cX, \mathbb{Z})$ to a ``modified Mukai vector'' as follows \cite{Alexandrov:2010ca}
\beq
H_3(\cX, \mathbb{Z})\, \ni \, \gamma \quad   \stackrel{\text{\tiny{mirror map}}}{\longleftrightarrow}\quad \gamma=e^{A}\, \text{ch}(\cE)\sqrt{\text{Td}(\hat{\cX})}\, \in\,  H_{\text{even}}(\hat{\cX}, \mathbb{Z}).
\eeq
This is a realization of the ``topological mirror map'' discussed in \cite{DouglasTopMirrorMap}. The electric charges $q_{\Lambda}=(q_0, q_{a})$ now correspond to D0-D2 branes, while $p^{\Lambda}=(p^{a}, p^0)$ are D4-D6 charges. After the mirror map, the central charge is given by 
$ Z(\gamma;t)=\int_{\hat{\cX}}e^{-\mathcal{J}}\text{ch}(\cE)\sqrt{\text{Td}(\hat{\cX})}$, where $\mathcal{J}=B+iJ$ is the complexified K\"ahler form on $\cM_K(\hat{\cX})$.

Compactification to three dimensions promotes the special K\"ahler moduli space $\cM_K(\hat{\cX})$ to a quaternion-K\"ahler manifold $\widehat{\cM_K}(\hat{\cX})$, whose semiflat metric is deformed by instanton effects arising from Euclidean D0-D2-D4-D6 branes wrapping $\hat{\cX}\times S^{1}$, together with gravitational Taub-NUT instantons. This setup can furthermore be lifted to M-theory on $\cX\times T^{2}$, revealing a hidden $SL(2,\mathbb{Z})$-symmetry corresponding to the mapping class group of the torus. By T-duality, these effects are also equivalent to D(-1)-D1-D3-D5 and NS5-instanton corrections in the hypermultiplet sector of type IIB on $\hat{\cX}$, and the $SL(2,\mathbb{Z})$-symmetry is then recognized as the familiar S-duality group. 

These considerations imply that by using a combination of mirror symmetry and T-duality, all of the moduli spaces and instanton effects discussed above are in fact equivalent:
\beq
\cM_{{}^{\text{H}}}^{{}_\text{IIA}}(\cX)\quad \stackrel{\text{\tiny{T-duality}}}{\longleftrightarrow}\quad \widehat{\cM_C}(\cX)\quad  \stackrel{\text{\tiny{mirror map}}}{\longleftrightarrow}\quad  \widehat{\cM_K}(\hat{\cX})\quad \stackrel{\text{\tiny{T-duality}}}{\longleftrightarrow}\quad \cM_{{}^{\text{H}}}^{{}_\text{IIB}}(\hat{\cX}).
\label{qmirrorsymmetry}
\eeq
Because of this unification of moduli spaces in three dimensions, it seems that the appropriate language for studying the spectrum of BPS states in type II Calabi-Yau compactifications is that of quaternion-K\"ahler geometry, rather than special K\"ahler geometry.

\section{Instanton Partition Functions from Automorphic Forms}
\label{instpart}
We now return to the instanton partition function $\sZ_{{}^{\text{tot}}} $ in (\ref{totalinstantoncontribution}), and assume the theory  exhibits a U-duality symmetry $G_3(\mathbb{Z})$. We will argue that the exact partition function should be given by an automorphic form $\sF$, whose Fourier expansion  reproduces the instanton effects in (\ref{totalinstantoncontribution}). 

What properties should we expect of $\sF$? Recall that the partition function must at least be invariant under the semidirect product
\beq
G^{J}(\mathbb{Z})\equiv G_4(\mathbb{Z})\ltimes N(\mathbb{Z}),
\eeq
which is known as a \emph{Jacobi group} in the mathematical literature.\footnote{The group $G^J$ is sometimes also called ``Fourier-Jacobi group''.} The black hole partition function $\sZ_{{}^{\text{BPS}}}$ in (\ref{BPSinst}) should also be invariant under $G^{J}(\mathbb{Z})$, while the NS5-wave function $\Psi_{k, \ell^{\Lambda}}$ in (\ref{vectorvaluedtheta}) transforms in the Schr\"odinger-Weil representation of $G^{J}(\mathbb{Z})$. 

A useful analogue of this situation, first pointed out in \cite{Gunaydin:2006bz}, is the case of classical holomorphic Siegel modular forms, attached to the metaplectic representation of $Sp(4;\mathbb{Z})$. Fourier expansion around one of the cusps in the Siegel upper half plane yields coefficients attached to the Schr\"odinger-Weil representation of the Jacobi subgroup $SL(2,\mathbb{Z})\ltimes   N(\mathbb{Z})\subset Sp(4;\mathbb{Z})$, namely the standard Jacobi theta series \cite{MR781735,MR1634977}. This is known as a \emph{Fourier-Jacobi expansion}. 

Inspired by this analogy, we shall now discuss the construction of an automorphic form $\sF$ such that the generalization of the Fourier-Jacobi expansion for $G_3(\mathbb{Z}) \supset G^{J}(\mathbb{Z})$ reproduces the expected non-perturbative contributions to the partition function $\sZ_{{}^{\text{tot}}} $. For the general case when $G^{J}(\mathbb{Z})\subset G_3(\mathbb{Z})$ the metaplectic representation is replaced by the so called \emph{minimal representation} of $G_3(\mathbb{Z})$. This suggests that the NS5-partition function (\ref{TaubNutInst}) has an interpretation as the Fourier-Jacobi expansion of an automorphic form attached to the minimal representation of $G_3(\mathbb{Z})$ \cite{Weissmann,Gunaydin:2006bz}. 

\subsection{Constructing Automorphic Forms}
\label{constructing}
We are interested in constructing a function $\sF$ on the double coset $G(\mathbb{Z})\backslash G/K$, 
where $G$ is a real Lie group, $K$ its maximal compact subgroup and $G(\mathbb{Z})\subset G$ a discrete subgroup. The construction of such functions relies heavily on the representation theory of $G$. In general, automorphic forms arise from irreducible components in the decomposition of the space $L^{2}\big(G(\mathbb{Z})\backslash G\big)$, which forms a unitary representation of $G$. An  \emph{automorphic representation} is a pair $(\pi, \rho)$, where $\pi$ is an irreducible subspace in the decomposition of $L^{2}\big(G(\mathbb{Z})\backslash G\big)$, and $\rho$ is its realization. In order to ensure that the associated automorphic form lives on $G/K$, we look for distinguished $K$-invariant vectors $\cW_0\in \pi$. Such objects are known as \emph{spherical (Whittaker) vectors}. 

To spell out the construction more explicitly, we need the following data: 
\begin{enumerate}
\item  a unitary (automorphic) representation $(\pi, \rho)$ of $G$ acting on some Hilbert space $\sH$,
\item  a spherical vector $\cW_0\in \sH$, 
\item a $G(\mathbb{Z})$-invariant \emph{distribution} $\cW_{\mathbb{Z}}\in \sH^{*}$. 
\end{enumerate}
Using the bilinear pairing $\left<\cdot |\cdot \right> \, :\, \sH\times \sH^{*}\rightarrow \mathbb{R}$, one may construct $\sF$ as the overlap \cite{Kazhdan:2001nx,Pioline:2003bk}
\beq
\sF(g)=\left<\cW_{\mathbb{Z}}|\rho(g)\cdot \cW_0\right>, 
\label{automorphicform}
\eeq
where $g\in G$. This is manifestly invariant under the left action of $\gamma\in G(\mathbb{Z})$ due to the invariance of $\cW_{\mathbb{Z}}$, and under the right action of $k\in K$ since $\cW_0$ is spherical, and hence defines a function on $G(\mathbb{Z})\backslash G/K$. More generally, one may consider a non-spherical Whittaker vector $\cW_{K}$ (``finite $K$-type'') transforming in some representation of $K$, for which the associated automorphic \emph{form} $\sF_K(g)$ is no longer a function, but rather a section of a vector bundle over $G/K$. 

There is a powerful method to construct the $G(\mathbb{Z})$-invariant distribution $\cW_\mathbb{Z}$ using $p$-adic number theory. One may compute $\cW_\mathbb{Z}$ locally for each prime number $p$, giving a $p$-adic spherical vector $\cW_p$ invariant under $G(\mathbb{Z}_p)\subset G(\mathbb{Q}_p)$, where $\mathbb{Q}_p$ is the field of $p$-adic numbers and $\mathbb{Z}_p$ the ring of $p$-adic integers. It is justified to refer to $\cW_p$ as spherical since the discrete group $G(\mathbb{Z}_p)$ is the maximal compact subgroup $K_p\subset G(\mathbb{Q}_p)$, due to the fact that the ring $\mathbb{Z}_p$ is compact in $\mathbb{Q}_p$. By convention, the prime at infinity $p=\infty$ corresponds to the real numbers $\mathbb{Q}_\infty =\mathbb{R}$. Thus, $\cW_\infty$ coincides with the standard $K$-invariant vector $\cW_0$. 

The distribution $\cW_\mathbb{Z}$ is recovered by taking the product over all primes, yielding  \cite{Kazhdan:2001nx,Pioline:2003bk}
\beq
\sF(g)=\sum_{x\in \mathbb{Q}^\kappa}\prod_{p\, \text{prime}} \rho(g)\cdot \cW_p(x),
\label{padicform}
\eeq
where the sum over rational vectors $x\in \mathbb{Q}^{\kappa}, \, \kappa\in \mathbb{Z}_+$, encodes the bilinear pairing $\left<\cdot |\cdot \right>$. The representation-theoretic meaning of the integer $\kappa$ will be explained momentarily. We will also see below how to evaluate (\ref{padicform}) for explicit examples.

\subsection{5-Grading and the Heisenberg Parabolic}
\label{5grading}

Any Lie algebra $\mathfrak{g}$ of rank $>1$, with Lie group $G$, exhibits a 5-grading (vector space direct sums)
\beq
\mathfrak{g}=\mathfrak{g}_{-2}\oplus \mathfrak{g}_{-1}\oplus \mathfrak{g}_{0}\oplus \mathfrak{g}_{1}\oplus \mathfrak{g}_{2},
\label{5grading}
\eeq
where the subspaces $\mathfrak{g}_{\pm2}$ are both one-dimensional and are generated by the highest root of $\mathfrak{g}$, and $\mathfrak{g}_{\pm 1}$ are $2(\nv+1)$-dimensional vector spaces, endowed with a natural (Kirillov-Kostant) symplectic form.  Physically, the parameter $\nv$ will play the same role as in section \ref{BHpartition}, i.e. corresponding to the number of vector multiplets. $\mathfrak{g}_{\pm 1}\oplus \mathfrak{g}_{\pm2}$ form Heisenberg subalgebras with centers given by the highest and lowest root spaces $\mathfrak{g}_{\pm}$.

The 5-grading can be understood as the decomposition of the adjoint representation of $\mathfrak{g}$ with respect to the subalgebra $\mathfrak{sl}(2,\mathbb{R})=\mathfrak{g}_{-2}\oplus \mathbb{R}h\oplus \mathfrak{g}_{2}$, where $h$ is the Cartan generator associated to the highest root, and the subscripts in (\ref{5grading}) indicate the grade under the action of $h$. The level zero subspace  decomposes further into $\mathfrak{g}_0=\mathfrak{r}\oplus \mathbb{R}h$, where $\mathfrak{r}$ is the commutant of $\mathfrak{sl}(2,\mathbb{R})$ inside $\mathfrak{g}$.  At the level of the Lie group $G$, the subspace $\mathfrak{g}_{0}\oplus \mathfrak{g}_{1}\oplus \mathfrak{g}_{2}$ generates a parabolic subgroup $P\subset G$, with Levi decomposition $P=LN$, where the ``unipotent radical'' $N$ Êis the Heisenberg group generated by $\mathfrak{g}_{1}\oplus \mathfrak{g}_{2}$, and $L$ is the Levi factor generated by $\mathfrak{g}_0$. The parabolic subgroup $P$ is known as the \emph{Heisenberg parabolic}. We denote by $Z$ the center of $N$. 

To understand the physical relevance of the grading (\ref{5grading}) let us take $\mathfrak{g}$ to be the Lie algebra of the purported symmetry group $G_3$. The commutant $\mathfrak{r}$ is then identified with the Lie algebra of the duality group $G_4$ in four dimensions. The Cartan generator $h$ corresponds to rescalings of the radius $e^{-\phi}$, and hence the decomposition (\ref{5grading}) mirrors the decompactification limit $e^{-\phi}\rightarrow\infty$, which makes the $G_4$-invariance manifest. The one-dimensional subspace $\mathfrak{g}_2$ corresponds to translations of the NUT-scalar $\sigma$, while $\mathfrak{g}_1$ generates shifts of the Wilson lines $(\zeta^{\Lambda}, \tilde\zeta_\Lambda)$. Thus, the unipotent radical $N$ represents the Heisenberg action on $(\zeta^{\Lambda}, \tilde\zeta_\Lambda, \sigma)$ discussed in section \ref{3dinstanton}. Finally, the generators of the opposite unipotent radical $\mathfrak{g}_{-2}\oplus \mathfrak{g}_{-1}$ are non-linearly realized and correspond to a Geroch-type solution generating symmetry \cite{kinnersley,Breitenlohner:1987dg}. 

\subsection{Induced Automorphic Representations}
\label{FourierRepresentations}

A convenient way to construct an automorphic representation $(\pi, \rho)$ is through so called parabolic induction. Consider a function $f(n)\in \mathbb{R}$, with $n\in P\backslash G$, where $P\subset G$ is a parabolic subgroup. The group $G$ acts on $n$ from the right, and a compensating transformation of $p\in P$ from the left ensures that $png\equiv n'$ is again an element of $P\backslash G$. A representation $\rho$ of $G$ is obtained by acting on functions $f\in L^2(P\backslash G)$ according to
\beq
\rho(g)\cdot f(n)=\delta_\rho(p)f(n'), 
\eeq
where $\delta_\rho(p)\in \mathbb{R}$ is called the infinitesimal character of $\rho$. In the mathematical literature such induced representations are commonly denoted by $\text{Ind}_P^{G}\delta_\rho\equiv I(P)$. By choosing a $K$-covariant vector $\cW_K\in \sH=L^2(P\backslash G)$, the representation $I(P)$ extends to an automorphic representation $(\pi, \rho)$ through the construction in (\ref{automorphicform}). Since $G$ is a non-compact group, the Hilbert space $\sH$ Êis infinite-dimensional. However, one may still associate a dimension to the representation $(\pi,\rho)$; this is the so called \emph{functional, or Gelfand-Kirillov, dimension} $\kappa=\text{dim}\, P\backslash G$.

Elements of $I(P)$ are functions (or, more generally, sections) of $\kappa$ variables, corresponding to the summation variables $x\in \mathbb{Q}^\kappa$ in (\ref{padicform}). However, the $p$-adic Whittaker vector $\cW_p$ has only support on $\mathbb{Z}_p\subset \mathbb{Q}_p$, and therefore, after taking the product over all primes, the sum has support on integers $x\in \mathbb{Z}^\kappa$ (see \cite{Kazhdan:2001nx,Pioline:2003bk}). These integers should be identified with the physical charges of the theory. Hence, physically the automorphic form $\sF$ should be attached to a representation $(\pi, \rho)$ whose functional dimension coincides with number of charges. 

An example of particular relevance for what follows is when the parabolic subgroup $P$ coincides with the Heisenberg parabolic. The resulting representation $I(P,s)=\text{Ind}_P^{G}\delta_\rho({s})$ belongs to the (degenerate) principal continuous series of $G$ and depends on $\text{rank}_{\mathbb{R}}\,  \mathfrak{g}$ complex parameters collectively denoted by $s$. The functional dimension of $I(P,s)$ is $\kappa=\text{dim}\, \mathfrak{g}_1+1$. The automorphic functions attached to the degenerate principal series are non-holomorphic Eisenstein series 
\beq
\sE(g;{s})=\sum_{\gamma\in P(\mathbb{Z})\backslash G(\mathbb{Z})} \delta_\rho(\gamma g; {s}).
\label{Eisenstein}
\eeq
When the Heisenberg parabolic $P$ coincides with the Borel subgroup $B\subset G$ then the representation $I(B,{s})$ corresponds to the generic (non-degenerate) principal series. This happens in particular for the special cases of $SU(2,1)$ and $SL(3,\mathbb{R})$ to be considered in sections \ref{rigid} and \ref{extended}.

Although the BPS partition function (\ref{canonicalpartitionfunction}) only depends on $2(\nv+1)$ electric and magnetic charges, the 3d theory also exhibits the NUT-charge $k$, thus providing a total of $2\nv+3$ charges $(q_\Lambda, p^{\Lambda}, k)$,  which is precisely the functional dimension of $I(P,s)$. A naive first guess is therefore that the complete partition function $\sZ_{{}^{\text{tot}}} $ in (\ref{totalinstantoncontribution})  should be captured by an automorphic function in the degenerate principal series. A characteristic property of the Eisenstein series $\sE(g;{s})$ is that they are eigenfunctions of the Laplacian on the coset space $G/K$, with eigenvalue given by the quadratic Casimir operator. Physically, this is precisely the constraint enforced by supersymmetry in $\cN=8$ theories \cite{Green:1998by,Obers:1999um}, although we shall see in section \ref{conclusions} that this constraint is modified for $\cN=2$ theories.

One can also obtain smaller (``unipotent'') automorphic representations by taking residues of degenerate principal Eisenstein series $\sE(g;{s})$ along singular loci in the complex ${s}$-parameter space \cite{MR1469105,MR1767400}. In this way one can in particular construct the \emph{minimal representation} $\pi_{\text{min}}$ of smallest functional dimension $\kappa=\nv+2$ \cite{MR0342049}. The minimal representation can be obtained by quantizing the smallest coadjoint orbit $\cO_{\text{min}}$, i.e. the $G$-orbit of any root in the Lie algebra $\mathfrak{g}$. It is natural to choose the lowest root, corresponding to the subspace $\mathfrak{g}_{-2}$ in (\ref{5grading}). The orbit $\cO_{\text{min}}$ is then generated by the $(2\nv+4)$-dimensional subspace $\mathbb{R}h\oplus \mathfrak{g}_1\oplus \mathfrak{g}_{2}$. Quantization amounts to choosing a Lagrangian splitting of $\mathfrak{g}_1$ and realizing the generators as differential operators acting on functions of  ``coordinates'' and ``momenta'' \cite{MR1159103,Kazhdan:2001nx,minspher}. The functional dimension of the associated representation $\pi_{\text{min}}$ is half the dimension of the orbit $\cO_{\text{min}}$, as advertised. Physically, one may think of these variables as the $\nv+1$ magnetic charges $p^{\Lambda}$ (or, in the opposite polarization, electric charges $q_\Lambda$) together with the NUT charge $k$ \cite{Pioline:2005vi}. Formally, $k$ therefore plays the role of a quantization parameter, by analogy with Planck's constant  $\hbar$. The minimal representation generalizes the metaplectic representation for symplectic groups, and its Fourier-Jacobi expansion yields theta series in the Schr\"odinger-Weil representation of $G^{J}(\mathbb{Z})$ \cite{Weissmann}, which is what we expect for the NS5-partition function (\ref{TaubNutInst}).

\subsection{Abelian Fourier Expansion}

In order to isolate the dependence on the scalars $(\zeta^{\Lambda}, \tilde\zeta_\Lambda, \sigma)$ as in the instanton sums (\ref{totalinstantoncontribution}) one must perform a Fourier expansion of $\sF$ which diagonalizes the action of $N(\mathbb{Z})$. However, since this group is non-abelian one cannot decompose with respect to all the generators simultaneously. We shall therefore first perform the decomposition along the center $Z$ of $N$. 

The Fourier expansion along $Z$ reads
\beq
\sF(g)=\sum_{\chi_Z\in \hat{Z}} \sF_{\chi_Z}(g),
\label{FourierCenter}
\eeq
where $\chi_Z\, :\, Z \, \rightarrow \, \mathbb{C}^{*}$ is a unitary character of $Z$ which is trivial on $Z(\mathbb{Z})$. In other words, the sum runs over the unitary irreducible representations of $Z$, known as the ``unitary dual'' $\hat{Z}$. The Fourier coefficient $\sF_{\chi_Z}(g)$ is formally defined through the integral
\beq
\sF_{\chi_Z}=\int_{Z(\mathbb{Z})\backslash Z(\mathbb{R})} \sF(zg) \overline{\chi_Z(z)} dz.
\eeq 
More explicitly, in our conventions we may take $\chi_{Z}(\sigma)=e^{-\pi i k\sigma}$ and write
\beq
\sF_k(t, \zeta, \tilde\zeta, \phi)=\int_{0}^{2} \sF(t, \zeta, \tilde\zeta, \sigma, \phi) e^{i\pi k \sigma} d\sigma.
\label{CenterCoefficient}
\eeq
We restrict first to the case when $Z$ acts trivially, corresponding to the zeroth Fourier coefficient $ \sF_0$. The abelian expansion of $\sF_{\text{A}}\equiv\sF_0$ with respect to ${N_{\text{a}}}=N/Z$ is given by $\sum_{\chi_{{N_{\text{a}}}}\in \hat{N}_{\text{a}}} \sF_{\chi_{{N_{\text{a}}}}}$, where the $\chi_{{N_{\text{a}}}}$:th Fourier coefficient is
\beq
\sF_{\chi_{{N_{\text{a}}}}}(g) =\int_{{N_{\text{a}}}(\mathbb{Z})\backslash {N_{\text{a}}}(\mathbb{R})} \sF(ng) \overline{\chi_{{N_{\text{a}}}}(n)} dn.
\label{abeliancoefficient}
\eeq

If we choose a Lagrangian splitting of $\mathfrak{g}_1$ such that the characters $\chi_{{N_{\text{a}}}}$ of  ${N_{\text{a}}}(\mathbb{R})$ takes the explicit form $\chi_{{N_{\text{a}}}}(\zeta, \tilde\zeta)=e^{-2\pi i (q_\Lambda \zeta^{\Lambda}-p^{\Lambda}\tilde\zeta_\Lambda)}$, where $(q_\Lambda, p^{\Lambda})\in \mathbb{Z}^{2\nv+2}$, then we can write the abelian Fourier expansion more explicitly as
\beq
\sF_{\text{A}}(t, \zeta, \tilde\zeta, \phi)=\sum_{(q_\Lambda, p^{\Lambda})\in \mathbb{Z}^{2n+2}}C(q_\Lambda, p^{\Lambda}) \cW_{q_\Lambda, p^{\Lambda}}(t,\phi) e^{-2\pi i (q_\Lambda \zeta^{\Lambda}-p^{\Lambda}\tilde\zeta_\Lambda)},
\label{abelianterm}
\eeq
where we have also extracted the purely numerical coefficients $C(q_\Lambda, p^{\Lambda})$.  The moduli dependent part $ \cW_{q_\Lambda, p^{\Lambda}}(t,\phi)$ corresponds to the spherical Whittaker vector $\cW_0$ in a ``Fourier transformed representation''. Similarly, the numerical Fourier coefficients $C(q_\Lambda, p^{\Lambda})$ are related to the $p$-adic spherical vector $\cW_p$, after performing a $p$-adic Fourier transform and evaluating the product over all primes (see \cite{Kazhdan:2001nx,Pioline:2003bk,Persson:2010ms} for details).

The result (\ref{abelianterm}) bears an strong resemblance with the instanton sum (\ref{BPSinst}). The instanton effects should be reproduced by the leading term in the expansion of the spherical vector $ \cW_{q_\Lambda, p^{\Lambda}}(t,\phi)$ around the cusp $e^{\phi}\rightarrow 0$. It is furthermore suggestive to identify the BPS-degeneracies $\overline{\Omega}(\gamma)$ with the numerical Fourier coefficients $C(q_\Lambda, p^{\Lambda})$. 

\subsection{Non-Abelian (Fourier-Jacobi) expansion}
We shall now consider the case when the center $Z$ of $N$ acts non-trivially, corresponding to the generic coefficients $\sF_k, k\neq 0,$ in (\ref{CenterCoefficient}). This provides a generalization of the Fourier-Jacobi expansion of Siegel modular forms as disussed in the beginning of section \ref{instpart}. 

For $k\neq 0$ we write the expansion (\ref{FourierCenter}) explicitly as 
\beq
\sF_{\text{NA}}(t, \zeta, \tilde\zeta, \sigma, \phi)=\sum_{k\neq 0} \sF_k(t, \zeta, \tilde\zeta, \phi) e^{-i\pi k \sigma}. 
\label{NAexpansion}
\eeq
When the center acts non-trivially one cannot simultaneously diagonalize all of the generators of the abelian subspace $\mathfrak{g}_1$, as was done in (\ref{abelianterm}). Instead, one must choose a \emph{polarization}, i.e. diagonalize either ``electric'' translations of $\zeta^{\Lambda}$ or ``magnetic'' translations of $\tilde\zeta_\Lambda$. 

For definiteness we consider a magnetic polarization, for which the general form of the non-abelian Fourier coefficients $ \sF_k, \, k\neq 0,$  is given by \cite{MR1726680,Pioline:2009qt,Bao:2009fg}\footnote{For details on the derivation, see for instance section 11.5 of \cite{Persson:2010ms}.}
\beq
 \sF_k(t, \zeta, \tilde\zeta, \phi)= \sum_{\ell^{\Lambda}\in \Gamma_{\text{m}}/(|k|\Gamma_{\text{m}})}\sum_{p^{\Lambda}\in \Gamma_{\text{m}}+\ell^{\Lambda}/|k|} \Psi_{k, \ell^{\Lambda}}(t, \phi, \zeta^{\Lambda}-p^{\Lambda})\, e^{2\pi i k p^{\Lambda}\tilde\zeta_\Lambda-\pi i k \zeta^{\Lambda}\tilde\zeta_\Lambda}.
\label{nonabelianFourier}
\eeq
The sum in (\ref{NAexpansion}) is then manifestly invariant under magnetic translations $\tilde\zeta_\Lambda \mapsto \tilde\zeta_\Lambda+n_\Lambda$, while electric shifts of the form $\zeta^{\Lambda}\mapsto \zeta^{\Lambda}+n^{\Lambda}$ require a compensating lattice shift $p^{\Lambda}\mapsto p^{\Lambda} +n^{\Lambda}$. Thus, the non-abelian contribution $\sF_{\text{NA}}$ is indeed invariant under the Jacobi group $G^{J}(\mathbb{Z})$. 

By comparing (\ref{nonabelianFourier}) with the analysis in section (\ref{NS5analysis}) it is clear that the non-abelian Fourier expansion corresponds to a sum over vector-valued theta series, with wave function $\Psi_{k, \ell^{\Lambda}}$. It is therefore suggestive to identify the Fourier coefficient $\sF_k$ with the partition function $(\ref{vectorvaluedtheta})$ for $k$ NS5-branes. This is consistent with the fact that $\Psi_{k, \ell^{\Lambda}}$ transforms in the Schr\"odinger-Weil representation of $G^{J}(\mathbb{Z})$. Moreover, $\sF_{\text{NA}}$ is a sum over $\nv+2$ variables $(p^{\Lambda}, k)$ which matches the functional dimension of the minimal representation of $G_3(\mathbb{Z})$. We conclude that (\ref{NAexpansion}) is the desired generalization of the Fourier-Jacobi expansion discussed in the beginning of this section.

%Summarizing our results so far, we deduce that the candidate automorphic partition function $\sZ_{{}^{\text{tot}}} $ decomposes as
%\beq
%\sZ_{{}^{\text{tot}}} (t,\zeta, \tilde\zeta, \sigma, \phi)= \sF_{\text{A}}(t,\zeta, \tilde\zeta, \phi)+ \sF_{\text{NA}}(t,\zeta, \tilde\zeta, \sigma, \phi), 
%\label{automorphiccorrections}
%\eeq
%mimicking the expected contributions from D- and NS5-brane instantons (\ref{totalinstantoncontribution}). In order to further test this proposal, we must find explicit forms of the wave functions $ \Psi_{k, \ell^{\Lambda}}(t, \phi, \zeta^{\Lambda})$ and in particular verify whether they exhibit the expected exponential suppression in the weak-coupling limit $e^{\phi}\rightarrow 0$. This will be done in the following sections for some examples. 

\section{Rigid Calabi-Yau Threefolds and the Universal Sector}
\label{rigid}
We will now put the abstract analysis of the previous section into our first explicit example. We consider a special class of $\cN=2$ theories, obtained through compactification of type IIA string theory on a \emph{rigid} Calabi-Yau threefold. A Calabi-Yau threefold is said to be rigid if it has no complex structure deformations; hence $h_{2,1}(\cX)=0$. We denote such a rigid threefold by $\cX_{\text{r}}$. 

The Hodge decomposition of $H^3(\cX_{\text{r}}, \mathbb{C})$ simplifies considerably:
\beq
H^3(\cX_{\text{r}}, \mathbb{C}) = H^{3,0}(\cX_{\text{r}}, \mathbb{C})\oplus H^{0,3}(\cX_{\text{r}}, \mathbb{C}). 
\label{rigidHodge}
\eeq
A consequence of this is that the intermediate Jacobian torus is an elliptic curve (see, e.g., \cite{MR2019160})
\beq
\cT(\cX)= H^{3}(\cX_{\text{r}}, \mathbb{R})/H^{3}(\cX_{\text{r}}, \mathbb{Z})=H^{3,0}(\cX_{\text{r}}, \mathbb{C})/H^{3}(\cX_{\text{r}}, \mathbb{Z})=\mathbb{C}/(\mathbb{Z}+\tau \mathbb{Z}), 
\eeq
where $\tau$ is the period variable which takes values in the upper half plane $SL(2,\mathbb{R})/U(1)$. Two elliptic curves $\mathbb{C}/(\mathbb{Z}+\tau\mathbb{Z})$ and $\mathbb{C}/(\mathbb{Z}+\tau'\mathbb{Z})$ are equivalent if $\tau'=\gamma\cdot \tau$ where $\gamma\in SL(2,\mathbb{Z})$. For simplicity, we further restrict our attention to those rigid Calabi-Yau threefolds for which the parameter $\tau$ takes values in the quadratic number field $\mathbb{Q}(\sqrt{-d})$, where $d$ is a square-free positive integer. Elliptic curves of this type are said to admit complex multiplication, or be of ``CM-type''. A rigid Calabi-Yau threefold is then of CM-type if and only if the associated Jacobian elliptic curve is of CM-type. The key observation is that in this case, the lattice $\mathbb{Z}+\tau\mathbb{Z}$ corresponds to the ring of integers $\mathcal{O}_d\subset \mathbb{Q}(\sqrt{-d})$. 

Physically, rigid Calabi-Yau compactifications are simpler, since the complex structure moduli space $\cM_C(\cX_{\text{r}})$ is trivial. For type IIB on $\cX_{\text{r}}$, the vector multiplet sector contains a single scalar modulus $X^0=\int_{\cA} \Omega_{3,0}$ and the prepotential is quadratic
\beq
F(X^{0})= \tfrac{1}{2}\tau (X^{0})^2.
\eeq
The modulus $X^0$ is non-dynamical and is customarily set to 1; however it is useful to retain it in order to keep track of the homogeneity degree of $F(X^{0})$. Since there are no complex structure moduli the BPS-index $\Omega(\gamma)$ is globally constant, and do not suffer wall-crossing. Equivalently, (\ref{rigidHodge}) ensures that one is automatically sitting at the end point of the attractor flow \cite{Moore:1998pn}. 

The vector multiplet sector in $D=4$ only contains the gravity multiplet, i.e. a single vector field $A_\mu$ (the graviphoton) together with the metric $g_{\mu\nu}$; this is the familiar Maxwell-Einstein theory. Upon dimensional reduction on $S^1$ it is well-known that this theory exhibits an enhanced symmetry group described by $U(2,1)$, and the classical moduli space in $D=3$ is given by the coset space \cite{kinnersley,Breitenlohner:1987dg}
\beq 
\cM_3^{\text{cl}}=U(2,1)/(U(2)\times U(1)).
\eeq
This 4-dimensional coset space is quaternion-K\"ahler\footnote{In fact, it is also K\"ahler, a curiosity among quaternion-K\"ahler manifolds.} and represents the classical part of the exact moduli space $\widehat{\cM_C}(\cX_{\text{r}})$ in $D=3$. It is parametrized by the radius $e^{-\phi}$, the NUT potential $\sigma$, and the two Wilson lines $(\zeta^0, \tilde\zeta_0)\equiv (\zeta, \tilde\zeta)$. Although $\cM_3^{\text{cl}}$ is described by a coset space, instanton corrections will deform the metric, and hence the exact moduli space $\cM_3^{\text{cl}}$ does not preserve this coset structure. 

Upon T-duality, $\cM_3^{\text{cl}}$ is mapped to the moduli space of the so called ``universal hypermultiplet'', parametrized by the dilaton $e^{\phi}$, the NS-axion $\sigma$ and the RR-scalars $(\zeta, \tilde\zeta)$ arising from the periods of the 3-form along the ``universal'' 3-cycles $(\cA, \cB)\in H_3(\cX_{\text{r}}, \mathbb{Z})$ \cite{Cecotti:1988qn,Ferrara:1989ik,Becker:1999pb}. 

Despite the fact that the quantum moduli space $\widehat{\cM_C}(\cX_{\text{r}})$ is not a coset, we shall assume that it retains an isometric action of an arithmetic subgroup $G_3(\mathbb{Z})\subset U(2,1)$. An analogous phenomenon is known to happen for the hypermultiplet sector in type IIB, where the moduli space $\cM_{{}^{\text{H}}}^{{}_\text{IIB}}({\cX_{\text{r}}})$ exhibits an isometric action of the S-duality group $SL(2,\mathbb{Z})$ \cite{RoblesLlana:2006is,Alexandrov:2009qq,Alexandrov:2010ca}. 

In \cite{Bao:2009fg,Bao:2010cc} it was proposed that for rigid Calabi-Yau threefolds of CM-type, the quantum theory should be invariant under the arithmetic U-duality group 
\beq
PU(2,1;\cO_d)\equiv U(2,1)\cap PGL(3,\cO_d), 
\eeq
known as the \emph{Picard modular group}. This group reproduces the correct Heisenberg shifts (\ref{Heis}), and is thus consistent with charge quantization For $d = 3\, \text{mod} \, 4$ the group $PU(2,1;\cO_d)$ induces a shift of $\sigma$ with non-trivial cocycle $\upsilon(m,n)=\tfrac{1}{2}(m+n +mn)$ in (\ref{Heis}) \cite{Bao:2010cc}, while for $d=1$, corresponding to the Gaussian integers $\cO_1=\mathbb{Z}+i\mathbb{Z}$, the cocycle is absent \cite{Bao:2009fg}. If taken literally, this has the interesting implication that for each rigid Calabi-Yau compactification there is a unique theta line bundle $\sL\rightarrow \cT(\cX_{\text{r}})$, and hence a unique NS5-partition function $\sZ_\Theta^{(k)}$, as was originally proposed by Witten \cite{Witten:1996hc}. 

To further test this proposal one would like to show that it is compatible with the expected form (\ref{totalinstantoncontribution}) of quantum corrections to the moduli space metric. Following the discussion in section \ref{FourierRepresentations}, we consider a $PU(2,1;\cO_d)$-invariant Eisenstein series in the principal series $I(B,s)$ of $U(2,1)$, which has the same functional dimension $\kappa=3$ as the number of D2-NS5 charges $(q, p, k)$.\footnote{This is moreover the most naive extension of previous successful results of summing up D(-1)-instantons in type IIB \cite{Green:1997tv,RoblesLlana:2006is} using an $SL(2,\mathbb{Z})$-invariant Eisenstein series attached to the principal series of $SL(2,\mathbb{R})$.} Using the techniques of section \ref{constructing} we obtain \cite{Bao:2009fg,Bao:2010cc}
\beq 
\sE^{{}_{PU(2,1;\cO_d)}}(g;s)=\sum_{(x_1, x_2)\in \mathbb{Q}(\sqrt{-d})^2\atop P(x_1, x_2)=0} \rho(g)\cdot \left[\prod_{p<\infty} \cW_p(x_1,x_2;s) \right] \cW_0(x_1,x_2;s),
\eeq
where $g$ is a representative in $ U(2,1)/(U(2)\times U(1))$ and $P(x_1, x_2)=0$ denotes a certain quadratic constraint on the summation variables (see \cite{Bao:2009fg,Bao:2010cc} for details), analogous to the lattice constraints discussed in \cite{Obers:1999um}. The principal series depends on a single parameter $s\in \mathbb{C}$, and the Eisenstein series converges absolutely for $\Re(s) > 2$. However, it is a famous result by Langlands \cite{MR0579181} that $\sE_{{}^{PU(2,1;\cO_d)}}(g;s)$ can be analytically continued to a meromorphic function in the full complex $s$-plane, away from the poles at $s=0, 2$. Indeed, $\sE^{{}_{PU(2,1;\cO_d)}}(g;s)$ satisfies a functional equation relating its value at $s$ to the Weyl-transformed value at $2-s$ \cite{Bao:2009fg,Bao:2010cc}.

The spherical vector $\cW_0$ is a function on $B\backslash U(2,1)$, and hence depends on three real variables, or equivalently on 2 variables $(x_1, x_2)\in \mathbb{Q}(\sqrt{-d})^2$ subject to a constraint $P(x_1, x_2)=0$. The explicit form of $\cW_0$ is obtained by taking the unitary norm of the first row of a representative $n\in P\backslash U(2,1)$; the norm ensures that $\cW_0$ is $U(2)\times U(1)$-invariant, i.e. spherical. The $p$-adic counterpart $\cW_p$ is similarly obtained by taking the corresponding $p$-adic norm $|z|_p^{\mathbb{Q}(\sqrt{-d})}\equiv \sqrt{|z\bar{z}|_p}, \, z\in \mathbb{Q}(\sqrt{-d})$. For the explicit formulas and more details on this construction, see App. A. of \cite{Bao:2009fg}, or references \cite{Kazhdan:2001nx,Pioline:2003bk,Persson:2010ms} for other examples.

By computing the Fourier expansion of $\sE^{{}_{PU(2,1;\cO_d)}}(g;s)$ one then obtains explicit candidates for the black hole partition function  in (\ref{abelianterm}) and the NS5 wave function $\Psi_{k, \ell}(\phi, \zeta)$ in (\ref{nonabelianFourier}). Extracting first the moduli-dependent part $\cW_{q, p}(\phi)$ of $\sE^{{}_{PU(2,1;\cO_d)}}_{q, p}(\phi)$ one finds
\beq
\cW_{q, p}(\phi;s)= K_{2s-2}\left(2\pi e^{-\phi}\frac{|q+\tau p|}{\sqrt{\Im(\tau)}}\right),
\label{WhittakerabelianU21}
\eeq
where $K_{2s-2}$ is the modified Bessel function and we recall that $\tau$ is the period variable of the elliptic curve $\cT(\cX)=\mathbb{C}/(\mathbb{Z}+\tau\mathbb{Z})$. For the special value $s=3/2$, the Whittaker vector $\cW_{q,p}(\phi;3/2)$ reproduces the known form of the D2-instanton series in type II on a Calabi-Yau threefold \cite{Alexandrov:2008gh}. Indeed, upon expanding $\cW_{q,p}(\phi;3/2)$ around the weak-coupling cusp $e^{\phi}\rightarrow 0$, one finds that the abelian Fourier expansion (\ref{abelianterm}) takes the form
\beq
\sE^{{}_{PU(2,1;\cO_d)}}_{\text{A}}(\zeta, \tilde\zeta, \phi) {\sim} \sum_{(p,q)\in \mathbb{Z}^2}C_{\text{A}}(p,q;3/2) \exp\left[-2\pi e^{-\phi} \frac{|q+\tau p|}{\sqrt{\Im\tau}} -2\pi i (q\zeta -p\tilde\zeta)\right].
\label{abelianexpansion}
\eeq
This is indeed of the same form as the BPS-instanton series (\ref{BPSinst}) after noticing that for a rigid Calabi-Yau threefold the D2-brane central charge is \cite{Bao:2009fg}
\beq
Z(\gamma)=\frac{q+\tau p}{\sqrt{\Im(\tau)}}, \qquad \gamma=q\cA+p\cB\in H_3(\cX_{\text{r}}, \mathbb{Z}). 
\eeq
Note in particular that (\ref{abelianexpansion}) has the expected suppression $e^{-1/g_s}$ for D-brane instantons.  

 The numerical coefficients in (\ref{abelianexpansion}) were evaluated explicitly in \cite{Bao:2009fg,Bao:2010cc} with the result
\beq
 C_{\text{A}}(\Lambda;s)=\sum_{\omega'\in \cO_d \atop \Lambda / \omega' \in \cO_d^{*}} \left|\frac{\Lambda}{\omega'}\right|^{2s-2}\sum_{z\in \cO_d\atop \Lambda/(z\omega')\in \cO_d^{*}} |z|^{4-4s}, 
 \eeq
 where $\omega$ is a primitive vector in $\cO_d$ and $\Lambda=(q+\tau p)/\Im(\tau)$. As a first consistency check we note that for purely electric or purely magnetic charges, i.e. when either $q=0$ or $p=0$, the Fourier coefficients $C_{\text{A}}(q,0;3/2)$ reproduce the sum over divisors $\sum_{d|q} d^{-2}$ which is known to provide the instanton measure for D(-1)-instantons \cite{Green:1997tv,RoblesLlana:2006is}. 
 
 Let us now turn to the non-abelian Fourier coefficients (\ref{nonabelianFourier}). The wave function $\Psi_{k,\ell}$ extracted from $\sE^{{}_{PU(2,1;\cO_d)}}_{\text{NA}}$ takes the form \cite{Bao:2009fg}
\beq
\Psi^{{}_{PU(2,1;\cO_d)}}_{k,\ell}(\phi, \tilde\zeta-p)=\sum_{r=0}^{\infty} C_{\text{NA}}(r, k, \ell;s)\,  e^{-\pi |k|(\tilde\zeta-p)^2}\, H_r\left(\sqrt{2\pi |k|}(\tilde\zeta-p)\right)\, W_{-r-\tfrac{1}{2}, s-1}\left(4\pi |k| e^{-2\phi}\right) , 
\eeq
where $H_r$ is a Hermite polynomial and $W_{r,s}$ is the Whittaker function. Inserting $s=3/2$ and expanding the Whittaker function for weak coupling yields
\beq
\Psi^{{}_{PU(2,1;\cO_d)}}_{k, \ell}(\phi, \tilde\zeta-p)\sim \sum_{r=0}^{\infty}  C_{\text{A}}(r, k, \ell;3/2)\, H_r\left(\sqrt{2\pi |k|}(\tilde\zeta-p)\right)\, e^{-2\pi |k| e^{-2\phi}-\pi |k|(\tilde\zeta-p)^2}, 
\label{nonabelianWhittakerU21}
\eeq
which is indeed suppressed by $e^{-1/g_s^2}$ as befits NS5-brane instantons.\footnote{To be more accurate, in the special case of the Gaussian integers $\cO_1=\mathbb{Z}+i\mathbb{Z}$ the periodicity of the NS-axion is slightly different from (\ref{nonabelianFourier}), forcing the charge $k$ to be a multiple of 4 \cite{Bao:2009fg}.} Moreover, for fixed value of $r$ the dependence of $\Psi_{k, \ell}$ on $(\tilde\zeta-p)$ matches the general structure of the NS5-brane partition function found in \cite{Alexandrov:2010ca}, where it was argued that the insertion of a power of $(\tilde\zeta-p)$ in the generalized theta series (\ref{vectorvaluedtheta}) was related to the insertion of $(J_3)^2$ in the second helicity supertrace (\ref{secondhelicity}), as is required in order to get a non-vanishing result for $\cN=2$ theories. Unfortunately, as of yet there is no explicit expression for the numerical non-abelian Fourier coefficients $C_{\text{NA}}(r, k, \ell;s)$.

\section{The Extended Universal Hypermultiplet}
\label{extended}

For generic Calabi-Yau compactifications there is in general no known candidate for the spectrum generating symmetry $G_3(\mathbb{Z})$. However, it was suggested in \cite{Pioline:2009qt} that the S-duality symmetry $SL(2,\mathbb{Z})$ of the hypermultiplet moduli space in type IIB extends to an isometric action of $SL(3,\mathbb{Z})$. The motivation behind this can be understood in the T-dual picture of the vector multiplet sector in type IIA on $\cX\times S^{1}$. As already mentioned, the associated moduli space $\widehat{\cM_K}(\cX)$ lifts to M-theory on $\cX\times T^2$, thus carrying an action of $SL(2,\mathbb{Z})$. However, it is well-known that upon toroidal compactification to $D=3$ the naive mapping class group $SL(n, \mathbb{Z})$ of an $n$-torus is enhanced to the ``Ehlers symmetry'' $SL(n+1, \mathbb{Z})$. This suggests that the moduli space $\widehat{\cM_K}(\cX)$ carries a hidden isometric action of $SL(3,\mathbb{Z})$. 

To make this explicit, one can identify a 5-dimensional subspace $SL(3,\mathbb{R})/SO(3) \subset \widehat{\cM_K}(\cX)$ parametrized by the volume modulus $t$ of $\cX$, the RR-scalars $(\zeta^{0}, \tilde\zeta_0)\equiv (\zeta, \tilde\zeta)$ which couple to the (D0, D6)-branes, and the gravity scalars $(e^{-\phi}, \sigma)$. In the T-dual picture the role of the modulus $t$ is unchanged, while $(\zeta, \tilde\zeta)$ couple to D(-1)-D5 instantons, $\sigma$ to NS5-instantons, while the radius is as usual identified with the 4d string coupling $g_s=e^{\phi}$. By analogy with the standard universal hypermultiplet discussed section \ref{rigid}, the subspace $\{t, \zeta, \tilde\zeta, \sigma, \phi\}$ is referred to as the ``extended universal hypermultiplet''. 

It was further argued in \cite{Pioline:2009qt} that in the weak coupling, large volume limit, the full quaternionic moduli space decomposes locally into a product:
\beq
\widehat{\cM_K}(\cX)\sim SL(3,\mathbb{R})/SO(3) \times \cR_K(\cX)\times (\mathbb{R}^{3})^{\otimes h_{1,1}(\cX)}, 
\eeq
where $\cR_K(\cX)$ is the vector multiplet moduli space of M-theory on $\cX$, of real dimension $h_{1,1}-1$. The action of $SL(3,\mathbb{R})$ on $\widehat{\cM_K}(\cX)$ leaves this factor invariant but acts linearly on the vector space $(\mathbb{R}^{3})^{\otimes h_{1,1}(\cX)}$.

By a similar philosophy as in Section \ref{rigid} one can now attempt to sum up the contributions from D(-1)-D5 and NS5-brane instantons to the metric on $\widehat{\cM_K}(\cX)$ using an $SL(3,\mathbb{Z})$-invariant Eisenstein series $\sE^{{}_{SL(3,\mathbb{Z})}}(g; s_1, s_2)$,  depending on two parameters which are fixed to the values $(s_1, s_2)=(3/2, 3/2)$. The Eisenstein series can be constructed using similar methods as in Section \ref{rigid} (see \cite{Pioline:2003bk, Pioline:2004xq,Persson:2010ms}), and may also be written as a Poincar\' e series 
\beq
\sE^{{}_{SL(3,\mathbb{Z})}}(t, \phi, \zeta, \tilde\zeta,\sigma;s_1,s_2)=\sum_{\gamma\in B(\mathbb{Z})\backslash SL(3,\mathbb{Z})} \left(\gamma\cdot t\right)^{s_1-s_2}\left(\gamma\cdot e^{-2\phi}\right)^{s_1+s_2},
\eeq
where $t^{s_1-s_2}e^{-2\phi(s_1+s_2)}\equiv \delta_\rho(p;s_1,s_2)$ is the infinitesimal character of the principal series $I(B,s_1,s_2)$, and $B$ is the Borel subgroup. Using old results by Vinogradov-Takhtajan \cite{MR527787} and Bump \cite{MR2611825}, one may extract the explicit form of the Fourier coefficients of $\sE^{{}_{SL(3,\mathbb{Z})}}(g; s_1, s_2)$.  After taking the weak-coupling limit, the abelian contribution is of the form 
\beq
\sE^{{}_{SL(3,\mathbb{Z})}}_{\text{A}}(t, \phi, \zeta, \tilde\zeta;s_1, s_2)\sim \sum_{(p,q)\in \mathbb{Z}^2}C(p,q;s_1, s_2)\exp\left[-2\pi e^{-\phi} |Z(p,q)| -2\pi i (q\zeta-p\tilde\zeta)\right].
\eeq
As a non-trivial consistency check we note that the first term in the exponential reproduces the central charge for D0-D6 BPS-bound states \cite{Dhar:1998ip}:
\beq
|Z(p,q)|=\left(t|p|^{2/3}+t^{-1} |q|^{2/3}\right)^{3/2}.
\eeq
 The numerical coefficients $C(p,q;s_1, s_2)$ are given by a generalized sum over divisors of the charges $(p,q)$ whose explicit form can be found in \cite{Pioline:2009qt}.
 
 The non-abelian Fourier coefficients $\sE_{\text{NA}}^{{}_{SL(3,\mathbb{Z})}}$ give rise to a wave function $\Psi_{k, \ell}$ whose weak-coupling expansion reads
 \beq
 \Psi_{k, \ell}^{{}_{SL(3,\mathbb{Z})}}(\phi, \tilde\zeta-p)\sim \sum_{\tfrac{(kn)^{3}}{d^2}\in \mathbb{Z}} C\left(d, \tfrac{(nk)^3}{d^2}; s_1, s_2\right) \exp\left[-\tfrac{2\pi |k|e^{-\phi}\big(e^{-2\phi}+t^3(\tilde\zeta-p)^2+tn^2\big)^{3/2}}{e^{-2\phi}+t^3(\tilde\zeta-p)^2}+ \tfrac{2\pi i kt^3n^3(\tilde\zeta-p)}{e^{-2\phi}+t^3(\tilde\zeta-p)^2}\right]
 \label{nonabeliansl3wavefunction}
 \eeq
where $d=\gcd(k, \ell)$. We note the interesting fact that the numerical Fourier coefficients in (\ref{nonabeliansl3wavefunction}) are equivalent to the abelian coefficients $C(p,q;s_1, s_2)$ with the magnetic charge $p$ replaced by $d$ and the electric charge $q$ is replaced by the combination $(nk)^3/d^2$, thus representing an ``induced'' electric charge. This meshes well with the recent results of \cite{Alexandrov:2010ca}, which revealed that the instanton measure for bound states of D(-1)-D1-D3 instantons with D5-NS5-instantons is given by the generalized Donaldson-Thomas invariants $\Omega(q_\Lambda, p^{\Lambda})$ with $p^{0}$ replaced by $\gcd(p,k)$, where $(p,0)$ represents pure D5-branes and $(0,k)$ represents pure NS5-branes. Moreover, the exponential in (\ref{nonabeliansl3wavefunction}) matches the general NS5-brane instanton action found in \cite{Alexandrov:2010ca}. 
\section{Supersymmetry, Representation Theory and Twistors}
\label{conclusions}

We shall now outline how to add the final piece of the puzzle, namely the proper implementation of the constraints from supersymmetry. The degree of supersymmetry imposes constraints on the charges $(q_\Lambda, p^{\Lambda}, k)$ as well as on the geometry of the moduli spaces. We now discuss these constraints in turn, focussing first on charges in sections \ref{chargeorbits}-\ref{minimal}, while treating the moduli space geometry in sections \ref{twistor} and \ref{qauto}.

\subsection{$\mathcal{N}=8$ Charge Orbits and Representation Theory}
\label{chargeorbits}
Although we are primarily interested in the $\cN=2$ case, it is again useful to first draw lessons from theories with a higher degree of supersymmetry. For $\cN=8$ supersymmetry in $D=4$ a generic electric-magnetic charge vector $\gamma=(q_\Lambda, p^{\Lambda})$ is valued in a 56-dimensional symplectic lattice $\Gamma$, invariant under $Sp(56;\mathbb{Z})$ \cite{Hull:1994ys}. These charges fit into the {\bf 56}-representation of $E_{7(7)}(\mathbb{Z})$. The entropy of a BPS black hole is  \cite{Kallosh:1996uy}
\beq
S_{{}^{\text{BH}}}^{{}_{\cN=8}}(\gamma) = \pi \sqrt{\cI_4(\gamma)},
\eeq
where $\cI_4(\gamma)$ is the quartic $E_{7(7)}(\mathbb{Z})$-invariant. BPS-solutions require $\cI_4\geq 0$. Generic $1/8$-BPS black holes satisfy $\cI_4\neq 0$, while states with $\cI_4=0$ but $\partial_\gamma \cI_4\neq 0$ correspond to $1/8$ BPS black holes with vanishing entropy \cite{Ferrara:1997ci}. States for which $\cI_4=\partial_\gamma\cI_4=0$ but $\partial_\gamma^2\cI_4 |_{\bf \text{Ad}}\neq 0$ preserve 1/4 supersymmetries, while states with $\partial_\gamma^2\cI_4 |_{\bf \text{Ad}}= 0$ preserve the maximal amount of 1/2 supersymmetries.\footnote{The expression $\partial_\gamma^2\cI_4 |_{\bf \text{Ad}}$ should be understood as the irreducible component of the Hessian of $\cI_4$, transforming in the adjoint representation of $E_{7(7)}$ (see \cite{Ferrara:1997ci,Ferrara:1997uz,Pioline:2006ni}). } 

The above conditions impose constraints on the BPS-index $\overline{\Omega}(\gamma)$, and consequently, via (\ref{Fouriergrowth}), also on the Fourier coefficients of the automorphic form $\sF$. This further translates into a choice of automorphic representation $(\pi,\rho)$ of the U-duality group $G_3(\mathbb{Z})$, since the number of charges determine the functional dimension $\kappa$ of $\pi$. 

As an example, we note that the constraint $\partial_\gamma^2\cI_4 |_{\bf \text{Ad}}= 0$ implies that 1/2 BPS-states in $\cN=8$ supergravity has support on 28 electric and magnetic charges. Combined with the NUT-charge $k$ this yields precisely the functional dimension $\kappa=29$ of the minimal representation $\pi_{\text{min}}$ of $E_{8(8)}(\mathbb{Z})$ \cite{Gunaydin:2001bt}. Indeed, it was conjectured in \cite{Pioline:2005vi} that the minimal theta series of $E_{8(8)}$ provides the partition function of 1/2 BPS-states in $\cN=8$ supergravity. This is also consistent with recent results revealing that the same theta series captures the non-perturbative contributions to the $E_{8(8)}(\mathbb{Z})$-invariant coefficient of the 1/2-BPS saturated $R^4$-coupling in type II string theory on $T^7$ \cite{Pioline:2010kb} (see also \cite{Green:2010kv}). Concretely this implies that when $\sF$ is in the minimal representation, the associated Fourier coefficients $C(q_{\Lambda}, p^{\Lambda})$ in (\ref{abelianterm}) vanish unless their arguments satisfy  $\partial^2\cI_4(q_\Lambda, p^{\Lambda}) |_{\bf \text{Ad}}= 0$. A simple example of this phenomenon is the minimal representation of $SL(3,\mathbb{R})$ for which the Fourier coefficients vanish unless $pq=0$ (see, e.g., \cite{Pioline:2009qt}).

% A perhaps more familiar analogue of this phenomenon occurs for the holomorphic discrete series of $SL(2,\mathbb{R})$, in which case the associated automorphic representations correspond to the standard weight $2r$ holomorphic Eisenstein series $\sum (m_1+m_2\tau)^{-2r}$, whose Fourier coefficients $a(n)$ vanish unless $n\geq 0$. 
 
 \subsection{Magic $\cN=2$ Supergravities and Quaternionic Representations}
 \label{magic}
 
 Let us now turn to $\cN=2$ theories. The general form of the Bekenstein-Hawking entropy is given by
 \beq
 \cS_{{}^{\text{BH}}}^{{}_{\cN=2}}(\gamma)=\pi \sqrt{\mathcal{Q}_4(\gamma)} 
 \eeq
 where $\mathcal{Q}_4(\gamma)$ is a quartic polynomial in the charges. For example, for 1/2 BPS black holes in M-theory on $\cX\times S^{1}$, corresponding to M2-branes wrapped on 2-cycles with $p^{0}=0$, the polynomial is given by $\mathcal{Q}_4=d_{ABC}p^{A}p^{B}p^{C}q_0$, $A=1, \dots, h_{1,1},$ and $d_{ABC}\sim \int J_A\wedge J_B\wedge J_C$, with $J_A\in H^{1,1}(\cX, \mathbb{Z})$ a generator of the K\"ahler cone \cite{Maldacena:1997de}. 
 
Whenever the $\cN=2$ theory exhibits a symmetry $G_4$ in $D=4$, the quartic polynomial $\mathcal{Q}_4(\gamma)$ is again identified with the associated quartic invariant $\cI_4(\gamma)$. An interesting class of toy models where this happens is provided by the \emph{magic} $\cN=2$ supergravity theories \cite{Gunaydin:1983rk,Gunaydin:1983bi}, which are classified by degree 3 Jordan algebras $J$, equipped with a cubic norm $N_3$ (see \cite{Gunaydin:2005gd} for a survey). The classical moduli spaces in $D=4$ are special K\"ahler coset spaces,
\beq
\cM^{\text{cl}}_4=\text{Conf}(J)/(U(1)\times M)
\eeq
 with cubic prepotentials of the form $F(X)=N_3(X)/X^{0}$. Here, $\text{Conf}(J)$ Êis the conformal group leaving the ``cubic lightcone'' $N_3=0$ invariant \cite{Gunaydin:2000xr}, and $K_4=U(1)\times M$ is its maximal compact subgroup. The factor $M$ is known as the ``reduced structure group'' of $J$. 

After compactification to $D=3$ the moduli space is enhanced to a quaternionic coset space 
\beq
\cM^{\text{cl}}_3=\text{QConf}(J)/(SU(2)\times \widetilde{\text{Conf}(J)},
\eeq 
where $\widetilde{\text{Conf}(J)}$Ê is a compact form of $\text{Conf}(J)$, and $\text{QConf}(J)$ is the quasi-conformal group of $J$ which leaves invariant a ``quartic lightcone'' $\cN_4=0$ inside the complex vector space $\mathfrak{g}_1^{\mathbb{C}}\oplus \mathfrak{g}_2^{\mathbb{C}}$. In terms of formal complexified ``charges'' $(q_\Lambda^{\, {}_{\mathbb{C}}}, p^{\Lambda}_{{}^{\mathbb{C}}}, k_{{}^{\mathbb{C}}})$ one can write the $\cN_4$ as \cite{Gunaydin:2007qq,Gunaydin:2006bz}
\beq 
\cN_4(q_\Lambda^{{}_{\mathbb{C}}}, p^{\Lambda}_{{}^{\mathbb{C}}}, k_{{}^{\mathbb{C}}})=\cI_4\left(q_\Lambda^{\, {}_{\mathbb{C}}}-\bar{q}_\Lambda^{\, {}_{\mathbb{C}}}, p^{\Lambda}_{{}^{\mathbb{C}}}-\bar{p}^{\Lambda}_{{}^{\mathbb{C}}}\right)+2\left(k_{{}^{\mathbb{C}}}-\bar{k}_{{}^{\mathbb{C}}}+p^{\Lambda}_{{}^{\mathbb{C}}}\bar{q}_\Lambda^{\, {}_{\mathbb{C}}}-\bar{p}^{\Lambda}_{{}^{\mathbb{C}}}q_\Lambda^{\, {}_{\mathbb{C}}}\right)^2,
\label{qlightcone}
\eeq
where $\cI_4$ is the quartic $\text{Conf}(J)$-invariant. This expression will play a role in section \ref{qauto} below.

 In general it is not known whether these magic theories arise from string compactifications. Nevertheless, it has been speculated that they have well-defined quantum completions which are invariant under discrete subgroups $\text{QConf}(J, \mathbb{Z})\subset \text{QConf}(J)$ (see for instance \cite{Gunaydin:2005gd,Gunaydin:2005mx,Gunaydin:2006bz,Gunaydin:2007qq}). An example of particular interest is when $J=\mathbb{R}$ and $\text{Conf}(J)=SL(2,\mathbb{R})$, in which case $\cM_3^{\text{cl}}=G_{2(2)}/SO(4)$. This exceptional coset space also arises as the classical vector multiplet moduli space in type IIA compactified on $\cX\times S^{1}$, with $\nv=h_{1,1}(\cX)=1$, or, equivalently, as the hypermultiplet moduli space in type IIB on $\cX$ \cite{Ferrara:1989ik,Bodner:1989cg}. 

For the quasiconformal groups $\text{QConf}(J)$ there exist a distinguished class of unitary representations leading to a similar classification of charges as was given above for $\cN=8$. This is the \emph{quaternionic discrete series} $\pi_\nu$ of Gross and Wallach \cite{MR1327538,MR1421947}. The representation $\pi_\nu$ depends on a single parameter $\nu$, and for $\nu\geq 2\nv+3$, it is irreducible and has functional dimension $\kappa=2\nv+3$, which is the same as for the principal series. In fact, the quaternionic discrete series can be embedded as a submodule in a degenerate principal series $I(P,s)$, for which the parameter $s$ is restricted to be integer. For smaller values of $\nu$, $\pi_\nu$ is no longer irreducible, but admits unitary irreducible sub-representations $\pi'_\nu$ of smaller functional dimension. For instance, when $\nu=(\nv+2)/3$ one obtains the quaternionic analogue of the minimal representation $\pi_{\text{min}}$, of functional dimension $\kappa=\nv+2$. Following \cite{MR1327538} we denote this minimal quaternionic representation by $\pi_Z$, with $Z$ referring to the locus $\partial_\gamma^2\cI_4 |_{\bf \text{Ad}}= 0$. 

The physical relevance of the quaternionic discrete series in magic $\cN=2$ theories has been pointed out in a series of works \cite{Gunaydin:2005mx,Pioline:2006ni,Neitzke:2007ke,Gunaydin:2007bg,Gunaydin:2007qq,Pioline:2009qt}. In particular, it was shown that the BPS Hilbert space $\cH^{{}_{\text{BPS}}}$ furnishes a unitary irreducible representation in the quaternionic discrete series of $\text{QConf}(J)$. Concretely, this means that physical ``BPS wave functions'' should have support on $2\nv+3$ charges $(q_\Lambda, p^{\Lambda}, k)$. Based on these obervations it is suggestive that the total instanton partition function $\sZ_{{}^{\text{tot}}} $ for magic supergravities should be attached to the quaternionic discrete series $\pi_\nu$ of the symmetry group $\text{QConf}(J)$, rather than the degenerate principal series $I(P,s)$, as was tentatively assumed in the analysis of sections \ref{rigid} and \ref{extended} following a naive extrapolation of known results in $\cN=8$ theories. These ideas have been advocated in various earlier works \cite{Gunaydin:2005mx,Pioline:2006ni,Gunaydin:2007bg,Gunaydin:2007qq,Pioline:2009qt,Bao:2009fg}, although the lack of sufficiently explicit results in the mathematical literature has hindered progress.

\subsection{Minimal Theta Series and NS5-Branes}
\label{minimal}

To provide some evidence for the above speculations let us now revisit the analysis of section \ref{extended} for the extended universal hypermultiplet. Although the group $SL(3,\mathbb{R})$ does not afford any quaternionic representations, the coset space $SL(3,\mathbb{R})/SO(3)$ embeds naturally into the quaternionic symmetric space $G_{2(2)}/SO(4)$ discussed above. Assuming that the associated Calabi-Yau compactification preserves a discrete subgroup $G_{2(2)}(\mathbb{Z})$, one could hope to sum up the complete series of D(-1)-D1-D3-D5 and NS5-instanton corrections into an automorphic form in the quaternionic discrete series of $G_{2(2)}$.

As a first step in this direction, it was noticed in \cite{Pioline:2009qt} that $ \Psi_{k, \ell}^{{}_{SL(3,\mathbb{Z})}}$ in fact corresponds to a restriction of the spherical vector $\cW_0$ in the minimal quaternionic representation $\pi_Z$ of $G_{2(2)}$ \cite{MR1253210,Gunaydin:2007qq}. It was then conjectured that the automorphic theta series attached to the minimal representation of any group $\text{QConf}(J)$ should capture the (vector-valued) non-abelian NS5-brane partition function (\ref{vectorvaluedtheta}):
\beq
\sZ^{(k, \ell^{\Lambda})}(g)=\left< \cW^{k, \ell^{\Lambda}}_{\mathbb{Z}}\Big|\rho_{Z}(g)\cdot \cW_0^{k, \ell^{\Lambda}}\right> .
\label{NS5mintheta}
\eeq
 Indeed, the wave function $\Psi_{k, \ell^{\Lambda}}(p^{\Lambda})$ depends on $\nv+2$ variables $(p^{\Lambda}, k)$, which is precisely the functional dimension of the minimal representation $\pi_Z$. 

The conjecture (\ref{NS5mintheta}) is also consistent with the observations in (\ref{NS5analysis}) that the wave function $\Psi_{1,0}(p^{\Lambda})$ should be identified with the topological string amplitude $\Psi_{\text{top}}$ on $\cX$. For magic supergravities, the holomorphic anomaly equations satisfied by $\Psi_{\text{top}}$ are equivalent to the operator equations associated with the Joseph ideal, and thus intimately connected with the minimal representation \cite{Gunaydin:2006bz}. For $k>1$ the minimal wave function $\Psi_{k,\ell^{\Lambda}}(p^{\Lambda})$ provides a candidate for the one-parameter extension of the topological string amplitude proposed in \cite{Gunaydin:2005mx,Gunaydin:2006bz}. 

Although these results are encouraging, the theta series in the minimal quaternionic representation can at most capture part of the quantum corrections to the moduli space metric, notably those states whose charges obey the constraint $\partial_\gamma^2\cI_4 |_{\bf \text{Ad}}= 0$. Physically, this translates into the fact that the associated Fourier coefficients have support on ``very small charges'', meaning that there are no independent D1-D(-1) charges, but only those induced by the D3-D5 states. To include the contributions from more general BPS states one must consider automorphic representations attached to the generic quaternionic discrete series $\pi_\nu$, where the Fourier coefficients have support on all charges with $\cI_4(\gamma)> 0$ \cite{MR1988198,MR1932327,MR2251588}. This is the quaternionic generalization of the familiar fact that the Fourier coefficients $\sum a(n)q^{n}$ of the holomorphic $SL(2,\mathbb{Z})$-Eisenstein series $E_{2r}(\tau)=\sum(m_1+m_2\tau)^{-2r}$ vanish unless $n\geq 0$.

%This also fits nicely with the results of \cite{Alexandrov:2010ca}, which revealed that the instanton measure for $k>1$ NS5-branes should be identified with the (higher rank) generalized Donaldson-Thomas invariants $\overline{\Omega}(\gamma)$. This suggests that at least for magic supergravities, these automorphic techniques can in principle provide a non-perturbative completion of the topological string. 

\subsection{Twistor Space and Deformations of Quaternionic Manifolds}
\label{twistor}
Having identified the quaternionic discrete series $\pi_\nu$ as the physically relevant automorphic representation for $\cN=2$ supergravity theories, the next step is to implement the $G_3(\mathbb{Z})$-symmetry  while preserving the quaternion-K\"ahler property of the metric on the moduli space. To this end we shall require some new technology.

A quaternion-K\"ahler manifold $\cM$ is conveniently described through its \emph{twistor space} $\cZ$ \cite{MR664330}:
\beq
\mathbb{P}^{1}\quad \longrightarrow \quad \cZ \quad \longrightarrow\quad  \cM 
\label{twistorfibration}
\eeq
where the fiber $\mathbb{P}^1$ corresponds to the $S^2$ worth of almost complex structures carried by $\cM$. If $\cM$ is real $4(\nv+1)$-dimensional, the twistor space $\cZ$ is a complex contact manifold of complex dimension $2\nv+3$. The contact structure is encoded in a one-form $\cX_\cZ$ such that the top form $\cX_\cZ \wedge (\text{d}\cX_\cZ)^{\nv+1}$ is nowhere vanishing. Infinitesimal deformations of $\cM$ can be realized as deformations of the complex contact structure of $\cZ$, which are classified by $H^{1}(\cZ, \cO(2))$ \cite{lebrun1988rtq,lebrun1994srp}. Compatibility with $\cN=2$ supersymmetry is therefore ensured if quantum corrections (\ref{totalinstantoncontribution}) to the metric on the moduli space $\cM$ are implemented in terms of deformations of the contact one-form $\cX_\cZ$ \cite{Alexandrov:2008nk,Alexandrov:2008gh}. This situation corresponds to a supergravity generalization of the analysis in \cite{Gaiotto:2008cd}, where quantum corrections to the hyperk\"ahler metric on the moduli space of $\cN=2$ field theory on $\mathbb{R}^{3}\times S^{1}$ was studied. 

Following \cite{Alexandrov:2008nk,Alexandrov:2008gh}, one introduces a set of $2\nv+3$ local Darboux coordinates $(\xi^{\Lambda}, \tilde\xi_\Lambda, \alpha)$ on some open patch of $\cZ$, such that infinitesimal deformations of the contact structure (``contactomorphisms'') are realized as holomorphic sections $H(\xi, \tilde\xi, \alpha)\in H^{1}(\cZ, \cO(2))$. The Darboux coordinates can formally be viewed as complexifications of the axions $(\zeta^{\Lambda}, \tilde\zeta_\Lambda, \sigma)$ and hence the $N(\mathbb{Z})$ Heisenberg action lifts to an isometric action on $\cZ$:
\beq
\xi^{\Lambda}\,  \rightarrow \,  \xi^{\Lambda}+m^{\Lambda}, \qquad \tilde\xi_\Lambda \, \rightarrow \tilde\xi_\Lambda+n_\Lambda, \qquad \alpha\,  \rightarrow \alpha+2r -n_\Lambda \xi^{\Lambda} + m^{\Lambda}\tilde\xi_\Lambda+2\upsilon(m,  n).
\eeq
The pair $(\xi^{\Lambda}, \tilde\xi_\Lambda)$ therefore parametrize a complex symplectic torus $\cT_{\mathbb{C}}=\Gamma^{*}\otimes_{\mathbb{Z}} \mathbb{C}$ ($\Gamma^{*}$ being the dual of the charge lattice $\Gamma$). This is the complex torus introduced by Kontsevich and Soibelman \cite{Gaiotto:2008cd,ks}. The additional coordinate $\alpha$ then parametrizes the fiber of a ``contact line bundle'' $\sL_{\mathbb{C}}\rightarrow \cT_{\mathbb{C}}$ \cite{Alexandrov:2010ca}. 

When NS5-brane effects are absent the continuous shift symmetry of $\alpha$ is unbroken and the general contactomorphism $H(\xi, \tilde\xi, \alpha)$ reduces to a symplectomorphism acting on $\cT_{\mathbb{C}}$, which takes the following form \cite{Alexandrov:2008gh,Gaiotto:2008cd}:
\beq
H(\xi, \tilde\xi)= \sum_{\gamma\in \Gamma} \Omega(t;\gamma) \text{Li}_2\left(\lambda(\gamma)e^{-2\pi i (q_\Lambda \xi^{\Lambda}-p^{\Lambda}\tilde\xi_\Lambda)}\right)=\sum_{\gamma\in \Gamma} \lambda(\gamma)\overline{\Omega}(t;\gamma) e^{-2\pi i (q_\Lambda \xi^{\Lambda}-p^{\Lambda}\tilde\xi_\Lambda)}.
\label{symplecto}
\eeq
This is the infinitesimal version of the Kontsevich-Soibelman symplectomorphism $\mathcal{K}_\gamma$ \cite{ks}. A key observation for what follows is that $H(\xi, \tilde\xi)$ can also be viewed as a twistor space realization of the BPS instanton partition function $\sZ_{{}^{\text{BPS}}}$ in (\ref{BPSinst}). We now turn to discuss the expected structure of the general contactomorphism $H(\xi, \tilde\xi, \alpha)$ in the presence of NS5-effects.
\subsection{Instanton Partition Function on Twistor Space}
\label{qauto}

Following the logic of section \ref{instpart}, we wish to construct holomorphic sections $H(\xi, \tilde\xi, \alpha)\in H^{1}(\cZ, \cO(2))$ which transform nicely under an arithmetic Lie group $G(\mathbb{Z})$, such that the resulting metric on the quaternionic-K\"ahler base $\cM$ is invariant. Some results for $SL(2,\mathbb{Z})$ have been obtained in \cite{Alexandrov:2009qq,Alexandrov:2010ca}, though a general prescription is lacking. 

From the point of representation theory, there is a  proposal for how to implement these ideas when the moduli space $\cM_3$ Êis a coset space $G/K$. In \cite{MR1421947} it was shown that the quaternionic discrete series $\pi_\nu$ can be realized as the action of $G$ on holomorphic sections on the twistor space $\cZ\, \rightarrow \, G/K$. The reason for this can be understood quite naturally by noting that $\cZ\cong P_{\mathbb{C}}\backslash G_{\mathbb{C}}$, i.e. it is the complexification of the coset $P\backslash G$ discussed in section \ref{FourierRepresentations}. Morally, this is the complexification of the axions $(\zeta^{\Lambda}, \tilde\zeta_\Lambda, \sigma)\rightarrow (\xi^{\Lambda}, \tilde\xi_\Lambda, \alpha)$ mentioned above. The quaternionic discrete series therefore corresponds to the holomorphic quasi-conformal action of $G$ on $\mathfrak{g}_1^{\mathbb{C}}\oplus \mathfrak{g}_2^{\mathbb{C}}$ \cite{Gunaydin:2007qq,Gunaydin:2009zz,Gunaydin:2009dq}. Moreover, in terms of the Darboux coordinates, the quartic lightcone (\ref{qlightcone}) provides a K\"ahler potential for the twistor space according to
\beq
K_{\cZ}(\xi, \tilde\xi, \alpha)=\text{log}\, \cN_4(\xi, \tilde\xi, \alpha),
\eeq
which makes the $G$-invariance of $\cZ$ manifest.

By identifying the twistor fiber as $\mathbb{P}^1=SU(2)/U(1)$, the fibration in (\ref{twistorfibration}) is such that the twistor space takes the form $\cZ=G/(M\times U(1))$. Analogous to the realization of holomorphic modular forms for $SL(2,\mathbb{Z})$ as holomorphic sections of a line bundle over $SL(2,\mathbb{R})/U(1)$, automorphic forms in the quaternionic discrete series of $G$ have a realization in terms of holomorphic sections of $G$-equivariant vector bundles over $\cZ$. Moreover, as a module, one may identify the representation $\pi_\nu$ with the sheaf cohomology group \cite{MR1421947}:
\beq
\pi_\nu = H^1(\cZ, \cO(-\nu)).
\eeq
In terms of the Darboux coordinates $(\xi^{\Lambda}, \tilde\xi_\Lambda, \alpha)\in \cZ$, automorphic forms in the quaternionic discrete series $\pi_\nu$ correspond to holomorphic sections $\sF(\xi, \tilde\xi,\alpha)\in H^1(\cZ, \cO(-\nu))$. This is tantalizingly close to what we are after in describing deformations of the complex contact structure of the twistor space $\cZ$ itself. 

We then propose that the total instanton partition function (\ref{totalinstantoncontribution}) can be lifted to a holomorphic section $\sZ_{{}^{\text{tot}}} (\xi, \tilde\xi, \alpha)\in H^{1}(\cZ, \cO(-\nu))$ attached to the quaternionic discrete series $\pi_\nu$ of $G_3$, whose Fourier-Jacobi expansion with respect to $G^{J}(\mathbb{Z})=G_4(\mathbb{Z})\ltimes N(\mathbb{Z})$ exhibits the general structure
\beqa
\sZ_{{}^{\text{tot}}} (\xi, \tilde\xi, \alpha)&=&\sum_{(q_\Lambda, p^{\Lambda})\in \Gamma}\lambda(q_\Lambda, p^{\Lambda}) \overline{\Omega}(q_\Lambda, p^{\Lambda}) e^{-2\pi i (q_\Lambda\xi^{\Lambda}-p^{\Lambda}\tilde\xi_\Lambda)}
\nn \\
& & + \sum_{k\neq 0} \sum_{\ell^{\Lambda}\in \Gamma_{\text{m}}/(|k|\Gamma_{\text{m}})} \sum_{p^{\Lambda}\in \Gamma_{\text{m}}+\ell^{\Lambda}/|k|} \Psi_{k, \ell^{\Lambda}}(\xi^{\Lambda}-p^{\Lambda}) \, e^{2 \pi i kp^{\Lambda}\tilde\xi_\Lambda-\pi i k (\alpha-\xi^{\Lambda}\tilde\xi_\Lambda)} .
\nn \\
\label{autoquaternionicdisc}
\eeqa
 While the abelian contribution in (\ref{autoquaternionicdisc}) corresponds to an infinitesimal symplectomorphism of $\mathcal{T}_{\mathbb{C}}$, the non-abelian term provides the generating function of a generic contactomorphism of the twistor space $\cZ$ \cite{Alexandrov:2008gh,Alexandrov:2010ca}. 

In the spirit of \cite{Neitzke:2007ke,Gunaydin:2007bg,Alexandrov:2010ca}, the partition function $\sZ_{{}^{\text{tot}}} (t,\zeta, \tilde\zeta, \sigma, \phi)$ on the base $\cM$ is defined by quaternionic Penrose transform of (\ref{autoquaternionicdisc}):
\beq
\sZ_{{}^{\text{tot}}} (t,\zeta, \tilde\zeta, \sigma, \phi)\equiv e^{-\phi}\oint_{\mathcal{C}} \frac{\text{d}{\bf z}}{2\pi i \, {\bf z}} \sZ_{{}^{\text{tot}}} \Big(\xi({\bf z}), \tilde\xi({\bf z}), \alpha({\bf z})\Big),
\eeq
where ${\bf z}$ is a coordinate on the twistor fiber $\mathbb{P}^1$, and $\mathcal{C}$ is a suitable contour. Using the explicit form of the Darboux coordinates $(\xi^{\Lambda}, \tilde\xi_\Lambda, \alpha)$ in terms of $(x, {\bf z})\in \cM\times \mathbb{P}^1$ found in \cite{Neitzke:2007ke}, one can verify that the Penrose transform of the abelian term in (\ref{autoquaternionicdisc}) reproduces the moduli dependence $K_1(2\pi e^{-\phi}|Z(t;\gamma)|)$ as expected on general grounds for D-brane instantons \cite{Alexandrov:2008gh}. Indeed, for rigid Calabi-Yau threefolds, one recovers the Whittaker vector $\cW_{q,p}(\phi; 3/2)$ in (\ref{WhittakerabelianU21}). The saddle-point approximation of the non-abelian term in (\ref{autoquaternionicdisc}) has furthermore been evaluated in \cite{Alexandrov:2010ca}, revealing a generalized theta series with insertions as in (\ref{nonabelianWhittakerU21}).

\section{Concluding Discussion}
In this note we have reviewed proposals for implementing quantum corrections to the moduli space metric in $\cN=2$ supergravities on $\mathbb{R}^3\times S^{1}$, consistently with supersymmetry and $G_3(\mathbb{Z})$-invariance. In particular, we provided arguments that the complete instanton partition function $\sZ_{{}^{\text{tot}}}$ is an automorphic form attached to the quaternionic discrete series of $G_3$.

In light of this it would be interesting to revisit the analysis of section \ref{rigid} for compactifications on a rigid Calabi-Yau manifold $\cX_{\text{r}}$. In this case the relevant twistor space is a complex 3-dimensional coset space of the form $\cZ=U(2,1)/(U(1)\times U(1))$. Some of the relevant representation theoretic ingredients for constructing $U(2,1;\cO_d)$-automorphic forms on $\cZ$ in the quaternionic discrete series, such as the K-finite vectors $\cW_K^{(\epsilon)}$, have been analyzed in \cite{MR1988198,Gunaydin:2007qq}. One would like to construct $\sZ_{{}^{\text{tot}}} (\xi, \tilde\xi, \alpha)$ explicitly and extract the Fourier coefficients $C(p,q)$ which should provide an explicit form for the generalized Donaldson-Thomas invariants of $\cX_{\text{r}}$. 

In order to clarify the relation between the automorphic form $\sF(\xi, \tilde\xi, \alpha)\in H^{1}(\cZ, \cO(-\nu))$ and the infinitesimal deformation $H(\xi, \tilde\xi, \alpha)\in H^{1}(\cZ, \cO(2))$, a possibly useful tool is the \emph{twistor transform} \cite{MR610183,Pioline:2006ni,Neitzke:2007ke,Gunaydin:2007qq}. This is an involutive integral operator $\mathbb{T}$ that takes elements in $H^{1}(\cZ, \cO(-m))$ to elements in $H^{1}(\cZ, \cO(m-4))$. Consider for instance the example of $G_{2(2)}/SO(4)$ associated to a type IIB compactification on $\cX$ with $\nv=h_{1,1}=1$. The functional dimension of the quaternionic discrete series of $G_{2(2)}$ is $\kappa=5$, corresponding to the number of D(-1)-D1-D3-D5-NS5 charges $(q_\Lambda, p^{\Lambda}, k)$. Ideally, one should then construct an automorphic form $\sF(\xi, \tilde\xi, \alpha)\in H^{1}(\cZ, \cO(-6))$, and take the twistor transform $\mathbb{T}(\sF)\in H^1(\cZ, \cO(2))$ as a candidate for the deformation $H(\xi, \tilde\xi, \alpha)$ of the complex contact structure of $\cZ=G_{2(2)}/(U(1)\times SU(2))$. We note that the Fourier coefficients of modular forms on $G_{2(2)}$ generating the quaternionic discrete series have been analyzed in the mathematical literature \cite{MR1767400,MR1932327}; it would be very interesting to work out the underlying physics along the lines suggested above.

Finally we offer some remarks on the issue of wall crossing. For the Fourier coefficients in (\ref{autoquaternionicdisc}) to be identified with the BPS-index (\ref{secondhelicity}), one must verify that they are compatible with known wall crossing behaviour \cite{ks,Gaiotto:2008cd}. In the absence of NS5-brane effects, $k=0$, this is guaranteed since the abelian terms (\ref{symplecto}) coincide with infinitesimal Kontsevich-Soibelman symplectomorphisms. However, in the generic case, $k\neq 0$, it is presently unknown how to exponentiate the generating function $H(\xi, \tilde\xi, \alpha)$ into a finite contactomorphism of the twistor space $\cZ$. The non-abelian wall crossing problem would then involve a generalization of the KS formula to an infinite product of such contactomorphisms. Although this formulation is presently out of reach, the recent results of \cite{Alexandrov:2010ca} provide a suggestive hint: since the NS5-charge $k$ can be formally identified with a quantization parameter $\hbar$, i.e. with the center of a Heisenberg algebra, one might speculate that these contactomorphisms would act on the quantum torus of \cite{ks}, and the finite version of (\ref{autoquaternionicdisc}) would then presumably involve a product of quantum dilogarithms \cite{APP}.

%As a last word, it should be confessed that in this last section we have indulged in some rather handwaving speculations, and it is possible that some of these will eventually be proven misguided. Nonetheless, we believe that there is by now sufficient evidence that some version of the above story should provide an accurate description of the physics of $\cN=2$ supergravity, and we hope to report back on these issues in the near future, at least in some simplified situations. 
\section*{Acknowledgments}
I am grateful to Sergei Alexandrov, Ling Bao, Axel Kleinschmidt, Bengt E.W. Nilsson and Boris Pioline for extremely stimulating collaborations over the last few years on the results presented herein. I also wish to thank Matthias Gaberdiel, Stefan Hohenegger, P\"ar Kurlberg, Jan Manschot and Stefan Vandoren for related discussions. A special thanks goes out to Sergei Alexandrov, Christoffer Petersson and Boris Pioline for very helpful comments on a previous draft. Finally, I thank Eugen Paal and Alexander Stolin for organizing a stimulating workshop in a fantastic location. My research is supported in part by the Swiss National Science Foundation.

%can be thought of as the induced representation $\text{Ind}_{B_{\mathbb{C}}}^{\text{QConf}(J;\mathbb{C})} \delta_\rho(\epsilon) $, that is to say, the action of $\text{QConf}(J;\mathbb{R})$ on holomorphic sections on the \emph{complexified} coset $B_{\mathbb{C}}\backslash \text{QConf}(J;\mathbb{C})$. 
%\newpage 
\section*{References}
%\bibliography{iopart-num}
\bibliography{AGMPproceedings}
\bibliographystyle{iopart-num}
%\bibliographystyle{iopams}

%\begin{thebibliography}{9}
%\bibitem{iopartnum} IOP Publishing is to grateful Mark A Caprio, Center for Theoretical Physics, Yale University, for permission to include the {\tt iopart-num} \BibTeX package (version 2.0, December 21, 2006) with  this documentation. Updates and new releases of {\tt iopart-num} can be found on \verb"www.ctan.org" (CTAN). 
%\end{thebibliography}

\end{document}